\documentclass[prd,twocolumn,floatfix,superscriptaddress,nofootinbib]{revtex4-1}
\usepackage{dcolumn,amsmath}
\usepackage{graphicx}
\usepackage{braket}
\usepackage{bm}
\usepackage{amssymb}
\usepackage{hyperref}
\usepackage{multirow}
\usepackage{footnote}
\usepackage{subcaption}
\usepackage{ragged2e}
\usepackage{xcolor}
\usepackage{pdfcomment}
\usepackage[justification=justified,labelfont=bf,singlelinecheck=off]{caption}
\usepackage{threeparttable,booktabs}  
\usepackage{subcaption,siunitx,booktabs}
\usepackage{caption,afterpage,tabularx}
\usepackage{diagbox}
\usepackage{mathrsfs}
\usepackage{physics}
\usepackage{cancel}
\usepackage{amsmath}
\usepackage{mathtools}

\usepackage[title]{appendix}

\newcommand{\JSU}{
School of Physics and Electronic Engineering,
\\Jiangsu University, Zhenjiang, 212013 Jiangsu, China\\
}

\newcommand{\MUST}{State Key Laboratory of Lunar and Planetary Sciences, 
\\Macau University of Science and Technology, 999078 Macao, China}

\newcommand{\NJU}{Institute of Science and Technology for Deep Space Exploration, Nanjing University, Suzhou Campus, Suzhou, 215163 Jiangsu, China}

\begin{document}

\title{Baryon number violation accompanied by CP-violation as a quantum tunneling effect induced by superfluid pairing interactions}
\author{Yongliang Hao}
\affiliation{\JSU}
\author{Dongdong Ni}
\email{ddni@nju.edu.cn}
\affiliation{\NJU}
\affiliation{\MUST}

\date{\today}

\begin{abstract}

In this work, we explore a new picture of baryon number ($\mathcal{B}$) violation inspired by the formal analogies between the Brout–Englert–Higgs (BEH) model and the Ginzburg-Landau (GL) model. A possible manifestation of this new picture could be the transition between a pair of neutrons and a pair of antineutrons (i.e. $nn \rightarrow \bar{n}\bar{n}$), which violates $\mathcal{B}$ by 4 units (i.e. $|\Delta\mathcal{B}|=4$). In the presence of the superfluid pairing interactions, two neutrons can form a Cooper pair and can be modeled by a semi-classical complex scalar field, which carries two units of $\mathcal{B}$. In the presence of the $\mathcal{B}$-violating terms, the system does not possess a continuous $U(1)$ symmetry but instead it respect a discrete $Z_2$ symmetry. Before the spontaneous breaking of the $Z_2$ symmetry, the ground state (vacuum) of the neutron Cooper field and that of the antineutron Cooper field should have degenerate energy levels. After the spontaneous breaking of the $Z_2$ symmetry, the degeneracy of the ground states would be removed and a domain wall that interpolates between the two inequavalent ground states can emerge. If the vacuum energy of the neutron Cooper field is higher than that of the antineutron Cooper field, the false vacuum ($nn$) would decay into the true vacuum ($\bar{n}\bar{n}$) through a quantum tunneling process across the domain wall. Therefore, the $nn \rightarrow \bar{n}\bar{n}$ process can be considered as a false vacuum decay through a quantum tunneling process induced by the superfluid pairing interactions. Both the $\mathcal{B}$-violating and CP-violating effects can be quite naturally accommodated in the $nn \rightarrow \bar{n}\bar{n}$ process. We have also shown explicitly that it is not possible to completely rotate the CP-violating phases away by a rephrasing transformation unless some conditions are satisfied. The $\mathcal{B}$-violating process accompanied by CP-violation would open a promising avenue for exploring new physics effects beyond the Standard Model (SM).

\end{abstract}

\maketitle

\section{Introduction\label{sec1}}
The discovery of the Higgs-like boson \cite{aad2012observation,chatrchyan2012observation,chatrchyan2013observation} is considered as one of the major achievements of the Standard Model (SM). However, the SM is still far from perfect because there remain many puzzles that cannot be well explained within the framework of the SM \cite{zyla2020review}. Among such puzziles, an intriguing one that deserves a special attention is the matter-antimatter asymmetry, which refers to the observed excess of matter over antimatter in our universe \cite{zyla2020review}. Baryon number ($\mathcal{B}$) violation and CP-violation (along with C-violation) are two of the important conditions suggested by Sakharov to explain the observed matter-antimatter asymmetry \cite{sakharov1967violation}. On the one hand, $\mathcal{B}$ is usually regarded as an accidentally conserved quantity \cite{cerdeno2020impact}, although $\mathcal{B}$-violation may occur through some non-perturbative processes in the framework of the SM \cite{hooft1976symmetry,hooft1976computation,arnold2013simplified,ellis2016search,kuzmin1985anomalous}. Yet, no signal for $\mathcal{B}$-violation has been observed so far \cite{dev2022searches}. 
On the other hand, CP-violation makes it possible to absolutely distinguish between matter and antimatter \cite{landau1957conservation,okubo1958decay} and the evidence for CP-violation has been reported in numerous meson systems \cite{christenson1964evidence,aubert2001observation,abe2001observation,aaij2013first,aaij2019observation}. Furthermore, $\mathcal{B}$-violation and CP-violation play a critical role in testing the SM and in constructing new physics models because they are usually implemented as an important feature in many new physics models \cite{zyla2020review}.

The $n$-$\bar{n}$ oscillation process, which violates baryon number ($\mathcal{B}$) by two units (i.e. $|\Delta\mathcal{B}|=2$), has attracted an enormous level of attention both theoretically and experimentally \cite{phillips2016neutron}. It is predicted that CP-violation can possibly occur in the $n$-$\bar{n}$ oscillation process \cite{hao2022neutron}. The $\mathcal{B}$-violating process accompanied by CP-violating (CPV) effects opens a promising avenue for exploring new physics effects beyond the SM. Therefore, it is instructive to explore such possibility. Neutrons can serve as a powerful and versatile platform where many interesting processes can occur \cite{snow2022searches}, making it possible to search for new physical phenomena in a smaller experiment, comparing with the ones at the LHC. The searches for the $n$-$\bar{n}$ oscillation have been performed in various mediums \cite{phillips2016neutron}, including bound states, field-free vacuum, and etc. In field-free vacuum, the lower limit on the $n$-$\bar{n}$ oscillation time presented by the Institut Laue-Langevin (ILL) experiment is approximately $0.86 \times 10^{8}$ s \cite{baldo1994new}. In bound states, the $n$-$\bar{n}$ oscillations have been searched for by various experiments, such as Irvine-Michigan-Brookhaven (IMB) \cite{jones1984search}, Kamiokande (KM) \cite{takita1986search}, Frejus \cite{berger1990search}, Soudan-2 (SD-2) \cite{chung2002search}, Sudbury Neutrino Observatory (SNO) \cite{aharmim2017search}, Super-Kamiokande (Super-K) \cite{abe2015search,abe2021neutron}, and etc. Up to date, no significant signal for the $n$-$\bar{n}$ oscillation has been found.

In contrast to charged particles, the experiments with neutrons face more difficulties in both the acceleration and the detection of neutrons, leading to more uncontrollable limitations with the currently available experimental techniques. Furthermore, the $\mathcal{B}$-violating effects can be resulted from high-dimensional operators and thus are greatly suppressed by the energy scale of new physics. Specifically, the $n$-$\bar{n}$ oscillation process can be mediated by the new scalar bosons through the interactions described by dimension-9 operators. The direct searches for new heavy bosons at the LHC shows that no significant evidence of such new scalar bosons beyond the SM has been found so far \cite{zyla2020review}, suggesting that the new physics energy scale tends to be so high that a direct laboratory detection might be inappropriate through the current experimental techniques. Therefore, there are still many mysteries of neutron that need to be revealed. To circumvent the above-mentioned limitations, an alternative option is to perform the experiments or observations in a neutron-rich environment, instead of the ones at high energies. By contrast, neutron stars are one of the densest objects in our universe and contain so many neutrons that a tiny $\mathcal{B}$-violating effect can be amplified, making it possible to search for such signals through astrophysical observations. Another attractive option is to focus on the $\mathcal{B}$-violating processes where a pair of neutrons transit into a pair of antineutrons (i.e. the $nn \rightarrow \bar{n}\bar{n}$ transition process) \cite{mohapatra2021affleck}, rather than confining ourselves to the $n$-$\bar{n}$ oscillation process where a single neutron converts to a single antineutron. Comparing with the latter process, the former process violates baryon number by 4 units (i.e. $|\Delta\mathcal{B}|=4$).

In the subsequent sections, we will point out that such a transition can be manifested as a quantum tunneling process accompanied by CP-violation induced by the superfluid pairing interactions. Ref. As \cite{mohapatra2021affleck} pointed out, the transition from a pair of neutrons into a pair of antineutron (i.e. the $nn \rightarrow \bar{n}\bar{n}$ transition process) may possibly occur in the framework of the Affleck–Dine (AD) mechanism \cite{affleck1985new,dine1995supersymmetry,dine1996baryogenesis} or its variants \cite{babichev2019affleck,lloyd2021minimal}, even though the corresponding transition probability may be extremely small \cite{mohapatra2021affleck}. The baryogenesis and $\mathcal{B}$-asymmetry could be explained by the evolution of a complex scalar field carrying a non-zero baryon number in the AD mechanism \cite{affleck1985new,dine1995supersymmetry,dine1996baryogenesis}. Furthermore, due to the attractive interactions, two neutrons can possibly pair up to form a superfluid state \cite{migdal1959superfluidity}, which can be modeled by a semi-classical complex Cooper field. From the theoretical aspects, the neutron Cooper field can be described by the Ginzburg-Landau (GL) model \cite{ginzburg1950on,ginzburg2009superconductivity}), which has some historical connections and formal analogies to the Brout–Englert–Higgs (BEH) model \cite{englert1964broken,higgs1964broken1,higgs1964broken2,guralnik1964global} and the Bardeen-Cooper-Schrieffer (BCS) model \cite{bardeen1957theory}. In addition to this, the neutron Cooper field also has a non-zero baryon number and thus share some similarities with the AD field. Motivated by such similarities, we will explore the potential realization of the transition between a pair of neutrons and a pair of antineutrons through a complex Cooper field that carries a non-zero baryon number ($\mathcal{B}$). By invoking the formal analogies among the BEH, BCS and GL models, we will argue that such a realization may be accomplished in the presence of superfluid pairing interactions in a nuclear-matter environment, such as the interior of neutron stars or heavy nuclei.

We organize our discussions as follows. To begin with, we review the superfluidity of neutrons and the possible mathematical models that describe the transition between a pair of neutron and a pair of antineutron. Next, based on such models, we estimate the effective mass of the neutron Cooper pair. After that, we explore the possibility of observing CP-violation in the $nn \rightarrow \bar{n}\bar{n}$ transition process. Then, we briefly review the quantum tunneling from false vacuum (metastable state) to true vacuum (stable state). Finally, we transfer our attention to the possible physical consequences of the false vacuum decay. Following Refs. \cite{berti2023observation,lagnese2023detecting}, a stable state can be defined as a true vacuum whereas a metastable sate can be defined as a false vacuum. In the following discussions, unless otherwise specified, we will refer to the stable and metastable states as the true and false vacua, respectively. Moreover, we will adopt the natural units (i.e. $c \equiv 1$, $\hbar \equiv 1$).

\section{The Model}

The superfluidity and superconductivity are predicted to exist in nuclear systems, such as finite nuclei \cite{brink2005nuclear}, neutron stars \cite{haensel2007neutron}, and etc. In finite nuclei, the attractive interactions between nucleons may lead to their pairing and the corresponding phenomenon is analogous to the Cooper pairs of electrons in superconductors \cite{bohr1958possible}. In neutron stars, various baryons may pair up to form superconducting or superfluid states \cite{haensel2007neutron}. The possible phenomena related to the existence of nuclear superfluidity in compact stars involve pulsar glitches \cite{anderson1975pulsar,haensel2007neutron}, neutron-star cooling \cite{yakovlev2004neutron,page2004minimal,page2006cooling}, and etc. Particularly, it is suggested that neutron superfluids may exist in neutron stars \cite{migdal1959superfluidity}. The neutron superfluid pairing interactions can mainly be classified as various patterns, such as s-wave spin-singlet-sate ($^1$S$_0$), p-wave spin-triplet state ($^3$P$_2$), and etc., depending on the spin-orbital configurations of the neutron Cooper pairs (see e.g. Ref. \cite{haensel2007neutron}). At present, the thermodynamic property and the chemical composition of the neutron-star matter depend highly on theoretical assumptions, largely due to the lack of direct experimental information on the interiors of the neutron star \cite{oertel2017equations}. Since based on current theories the thermodynamic property and the chemical composition of the matter in neutron stars may vary greatly from the outer crust to the inner core, the internal structure of neutron stars can be divided into several internal layers or regions \cite{haensel2007neutron,potekhin2015neutron}. The patterns of neutron pairing would be different from region to region. It is expected that the neutron superfluidity in the crust region of neutron star can be attributed to the s-wave spin-singlet-sate ($^1$S$_0$) pairing \cite{wolf1966some}, while in the core region it can be attributed to the p-wave spin-triplet state ($^3$P$_2$) \cite{ruderman1967states,hoffberg1970anisotropic}.

Superfluidity and superconductivity are the low-temperature quantum effects. Neutron stars are initially hot after their born in supernova explosions and can cool down as a consequence of neutrino or photon emission \cite{haensel2007neutron}. When the temperature decreases below a critical temperature $T_c$, a superfluid phase transition may occur. Highly magnetized neutron stars, which are also known as magnetars, possess remarkably powerful magnetic fields, which are typically as strong as $10^{13}$-$10^{15}$ G near the surface and expected even stronger in the vicinity of the core \cite{mereghetti2015magnetars}. It is believed that the Cooper pairs can only exist in the presence of magnetic fields that are lower than a critical value $H_c$, which depends on the temperature \cite{mangin2017superconductivity} and could vary with environmental conditions.

Although neutrons, in a very brief picture, contain charged particles (e.g. quarks), they seem electrically neutral as a whole to the outside with very high precision \cite{zyla2020review}. In reality, neutrons can interact with the electromagnetic fields through the anomalous magnetic moment \cite{bjorken1964relativistic,bentez1990solution,*benitez1990erratum}. The spins of the paired neutrons in a s-wave spin-singlet state point to the opposite directions and tend to cancel out each other, making such paired neutrons effectively diamagnetic. In this case, it is reasonable to assume that the magnetic fields do not play an important role in the process under discussion (i.e. the $nn \rightarrow \bar{n}\bar{n}$ transition process). Therefore, to simplify our analysis, we focus our attention on the s-wave spin-singlet-sate. Furthermore, the s-wave spin-singlet neutron Cooper pair can be modeled by a complex scalar field. To be specific, in a semi-classical picture \cite{saxena2012high}, the scalar field and its charge conjugation can be identified as a pair of neutrons (i.e. $\phi \mapsto n_{\uparrow}n_{\downarrow}$) and a pair of antineutrons (i.e. $\phi^{\dag} \mapsto \overline{n}_{\uparrow} \overline{n}_{\downarrow}$), respectively. In summary, the neutron matter above the critical temperature can be identified as a normal state and the scalar field tends to vanish \cite{annett2004superconductivity}. The neutron matter below the critical temperature can be identified as a superfluid state and the scalar field tends to be different from zero \cite{annett2004superconductivity}.

In our analysis, the CP transformation is defined as (see e.g. Ref. \cite{mcdonald1995cosmological}):
\begin{equation}
\phi(t, \mathbf{x}) \xmapsto[]{\text{CP}} e^{i\delta} \phi^{\dag}(t, -\mathbf{x}).
\end{equation}
Here, $e^{i\delta}$ is an arbitrary phase factor and can be rotated away through a unitary transformation without causing any physical consequences \cite{greiner1996field}. Therefore, such a phase factor can be considered as a $U(1)$ rephrasing transformation. For simplicity of notation, we will also omit the time and space coordinates. In general, the scalar field can be decomposed in a linear representation:
\begin{equation}
\phi \equiv  \frac{1}{\sqrt{2}} \big(v+ \phi_1 + i \phi_2\big),
\label{phiv}
\end{equation}
or in an exponential (or polar) representation:
\begin{equation}
\phi \equiv  \frac{1}{\sqrt{2}} \big(v+ \phi_3\big) e^{i\frac{\phi_4}{v}},
\label{phiv}
\end{equation}
Here, $v$ is the vacuum expectation value (VEV) of $\phi$. In the linear representation, the fields $\phi_1$ and $\phi_2$ represent the real and imaginary components of $\phi$, respectively. For the question under discussion, we are allowed to define the real component as a CP-even field and define the imaginary component as a CP-odd field. In the exponential representation, the field $\phi_3$ is the radial component of $\phi$ and the field $\phi_4$ is the would-be Goldstone field. Both representations are expected to give equivalent results\footnote{Comparing with the linear representation, the exponential representation does not show an explicit renormalizability \cite{quevillon2022axion}. It could be more convenient to adopt the exponential representation in some cases, e.g. in the discussion of symmetry-breaking effects.}. In the following discussion, unless otherwise specified, we will adopt the linear representation by default but will switch to the exponential representation wherever convenient.

The Lagrangian density for the neutron superfluid in the absence of the electromagnetic fields can be given by
\begin{equation}
\mathscr{L}  \supset  \partial_{\mu} \phi^{\dag} \partial^{\mu} \phi - V(\phi^{\dag},\phi).
\end{equation}
Here, the two terms are associated with the kinetic energy (density) and scalar potential energy (density), respectively. In the absence of $\mathcal{B}$-violating terms, the scalar potential energy density $V$ can be replaced by the one in the GL model \cite{annett2004superconductivity,coleman2015introduction}:
\begin{equation}
V_1(\phi^{\dag},\phi) \equiv V_0 + \lambda_0 (\phi^{\dag}\phi) + \lambda_1 (\phi^{\dag}\phi)^2,
\label{landau}
\end{equation}
where the non-renormalizable high-order terms are omitted. The parameter $V_0$ is a constant associated with the normal state. For the neutron superfluid, the parameter $\lambda_0$ is related to the neutron mass $m_n$ by $\lambda_0 \simeq (2 m_n)^2$. The parameters $\lambda_0$ and $\lambda_1$ are two real parameters and may be functions of the temperature $T$, i.e. $\lambda_0 \equiv \lambda_0(T)$, $\lambda_1 \equiv \lambda_1(T)$. In order to have a superfluid phase transition, the parameter $\lambda_0$ should take a negative value (i.e. $\lambda_0<0$). A simple form of $\lambda_0(T)$ may be given by \cite{tinkham1996introduction}
\begin{equation}
\lambda_0(T) \equiv \lambda_0^{\prime} \Big( \frac{T}{T_c} - 1 \Big).
\end{equation}
Here, $\lambda_0^{\prime}$ is a real positive and temperature-independent parameter. When the temperature $T$ is higher than the critical temperature $T_c$, all the neutrons in the system would be disorganized and tend to be randomly aligned in space. In this case, the system described by Eq. (\ref{landau}) has a ground state (i.e. a vacuum state) at $|\phi| =0$ and can be referred to be in a normal state but not in a superfluid state. Since there is no preferred direction, the system possesses a $SO(3)$ symmetry. When the temperature $T$ is lower than the critical temperature $T_c$, there would be a superfluid phase transition, where two neutrons are combined into a Cooper pair. After the phase transition, the system has infinite degenerate ground states related by a $U(1)$ transformation at a non-zero $|\phi|\equiv v_0/\sqrt{2} =\sqrt{-\lambda_0/(2\lambda_1)}$. Furthermore, if the $U(1)$ transformation is a global phase transformation, the system would have a conserved quantity, which can be identified as the baryon number $\mathcal{B}$.

In the models with the scalar fields, some small $\mathcal{B}$-violating terms, such as the quadratic ($\phi^2$) \cite{babichev2019affleck,lloyd2021minimal,mohapatra2021affleck,mohapatra2022baryogenesis}, quartic ($\phi^4$) \cite{cline2020affleck,cline2020little}, or both the quadratic and quartic terms \cite{lloyd2023leptogenesis}, can be employed to generate the $\mathcal{B}$-asymmetry. Motivated by such models \cite{babichev2019affleck,cline2020affleck,cline2020little,lloyd2021minimal,mohapatra2021affleck,mohapatra2022baryogenesis,lloyd2023leptogenesis}, we choose the form of the scalar potential at the tree level as follows
\begin{equation}
\begin{split}
V(\phi^{\dag},\phi) \equiv & V_1(\phi^{\dag},\phi) + V_2(\phi^{\dag},\phi)\\
=& \lambda_0 (\phi^{\dag}\phi) + \lambda_1 (\phi^{\dag}\phi)^2 \\
&+ \Big(\lambda_2 e^{i \alpha} \phi^2 +  \lambda_3 e^{i \beta}  \phi^4 + \text{H.c.}\Big).\\
\end{split}
\label{eqvv}
\end{equation}
Here, $V_1$ and $V_2$ represent the $\mathcal{B}$-conserving and $\mathcal{B}$-violating terms, respectively. All the non-renormalizable terms with higher power of $\phi$ are assumed to be absent. Moreover, the linear ($\phi$) and cubic ($\phi^3$) terms can also be assumed to be excluded by imposing a discrete $Z_2$ symmetry: $\phi \mapsto -\phi$ \cite{lloyd2023leptogenesis}. The coupling parameters $\lambda_2$ and $\lambda_3$ are real numbers. Following Refs. \cite{mcdonald1995cosmological,haber2012group,branco2016group}, the non-trivial phase factors ($e^{i \alpha}$ and $e^{i \beta}$) can be added into the coupling parameters. The phases $\alpha$ and $\beta$ are also assumed to be real numbers. The presence of the non-trivial phase factors in the coupling parameters can be reasonable if some auxiliary fields are assumed to be present \cite{haber2012group}. For example, such phase factors can be originated from the vacuum expectation values of the spurion fields \cite{haber2012group}. We also assume that the parameters ($\lambda_2$, and $\lambda_3$) associated with the $\mathcal{B}$-asymmetry is small so that the $U(1)$ symmetry is still an approximate symmetry. In the absence of $V_2$ (i.e. $\lambda_2=0$, $\lambda_3=0$), the $U(1)$ symmetry and the $\mathcal{B}$-conservation can be restored. In the presence of $V_2$, the $U(1)$ symmetry is breaking and the additional phase factors would induce non-trivial effects, such as the $nn \rightarrow \bar{n}\bar{n}$ transition process.

\section{CP-violation and $\mathcal{B}$-violation \label{sec3}}

In this section, we will explore the compatibility conditions required by the coexistence of both $\mathcal{B}$-violation and CP-violation in the system of the neutron superfluid described by the scalar potential given by Eq. (\ref{eqvv}). We will argue that both effects can be quite naturally accommodated in the transition between a pair of neutrons and a pair of antineutrons  (i.e. $nn \rightarrow \bar{n}\bar{n}$) in the presence of the superfluid pairing interactions. We will focus on the origin and signatures of CP-violation associated with the scalar potential. As mentioned in Sec. \ref{sec1}, CP-violation can be established through the complex phase factors ($e^{i\alpha}$ and $e^{i\beta}$) of the coupling parameters and/or through the complex phase factors ($e^{i\gamma}$) of the vacuum expectation values \cite{branco1999cp,sozzi2007discrete,bigi2009cp}. The former is called explicit CP-violation while the latter is called spontaneous CP-violation \cite{branco1999cp,sozzi2007discrete,bigi2009cp}. For convenience of explanation, we will call $\alpha$ and $\beta$ the explicit CPV phases and call $\gamma$ the spontaneous CPV phase, though they do not necessarily lead to observable CPV effects.

\subsection{Spontaneous CP-violation \label{subsec1}}

The spontaneous CP-violation has been intensively studied over the past years (see e.g. Refs \cite{haber2012group,branco2016group}). Demonstrating the existence of CP-violation in an explicit way has never been a trivial task, as there are many subtleties and complexities that can easily be overlooked. First, we will review the main results in the literature (see e.g. Refs \cite{haber2012group,branco2016group}), because such information may help us to understand the intricate complexities of CP-violation better. Furthermore, previous studies pay more attention to the possibility of having the spontaneous CP violation, but focus less on the physical consequences of the spontaneous CP violation.

Following Ref. \cite{haber2012group}, we will focus on the physical consequences arising from the spontaneous CPV phase $\gamma$. We will show that the phase $\gamma$ would not lead to the $nn \rightarrow \bar{n}\bar{n}$ transition process. Here, we assume that the explicit CPV phases are absent (i.e. $\alpha=0$ and $\beta=0$). The scalar potential evaluated at $\langle \phi \rangle \equiv e^{i \gamma}v/\sqrt{2}$ satisfies the following expression:
\begin{equation}
\begin{split}
V\big\vert_{\phi =\langle \phi \rangle} =& \frac{1}{2}\lambda_0 v^2 + \frac{1}{4}\lambda_1 v^4 +  \lambda_2v^2 \cos(2\gamma) \\
&+ \frac{1}{2} \lambda_3v^4 \cos(4\gamma).    
\end{split}
\label{vcos}
\end{equation}
The minimum of the scalar potential can be obtained by differentiating the above expression with respect to the phase $\gamma$:
\begin{align}
&\begin{aligned}
\frac{\partial V}{\partial \gamma}\Bigg\vert_{\phi = \langle \phi \rangle}=\lambda_2 \sin(2\gamma)+ \lambda_3 v^2 \sin(4\gamma) = 0. 
\label{dvdgamma}
\end{aligned}
\end{align}
The solution to the above equation can be given by \cite{haber2012group}
\begin{equation}
\cos{(2\gamma)}\big\vert_{\text{min}} = -\frac{\lambda_2}{\lambda_3 v^2},
\label{cosmin}
\end{equation}
Apparently, the coupling parameters should satisfy the condition: $-1 \le -\lambda_2/(\lambda_3 v^2) \le 1$. We use the symbol $\gamma_m$ to denote the value of the phase $\gamma$ that satisfies the expression in Eq. (\ref{cosmin}). Spontaneous CP-violation requires that $\gamma_m$ is different from zero, i.e. $\gamma_m \neq k\pi$ ($k \in \mathbb{N}$). Therefore, unless $\gamma_m =k\pi$, spontaneous CP-violation would be established. However, such spontaneous CP-violation would not necessarily lead to observable CP-violating effects and the phase $\gamma_m$ would not be physical unless some additional conditions are satisfied (see e.g. Ref. \cite{branco1999cp}). Moreover, we could move a step back and analyze the possible role of the spontaneous CPV phase $\gamma$. Under CP transformation, the spontaneous CPV phase associated with the antineutron Cooper field ($-\gamma$) would differ from the one associated with the neutron Cooper field ($\gamma$) by a minus sign. Eq. (\ref{vcos}) shows that the minimum of the scalar potential is invariant with respect to the sign-flipping of $\gamma$. In other words, the minimum of the scalar potential is symmetric under the CP transformation (i.e. $V^{\text{CP}}=V$). In this case, the vacuum of the neutron Cooper field shares the same energy with that of the antineutron Cooper field. The tunneling from the neutron superfluid to the antineutron superfluid tends to cancel out the tunneling to the opposite direction. Therefore, such spontaneous CP-violation would play a trivial role in the $nn \rightarrow \bar{n}\bar{n}$ transition process.

\begin{figure*}[t]
\centering
\begin{subfigure}[h]{0.49\textwidth}
\centering
\includegraphics[width=\textwidth]{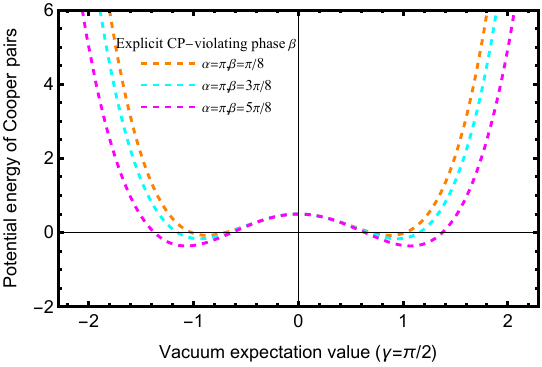}
\caption{Symmetric double well}
\label{fig1a}
\end{subfigure}
\hfill
\centering
\begin{subfigure}[h]{0.49\textwidth}
\centering         
\includegraphics[width=\textwidth]{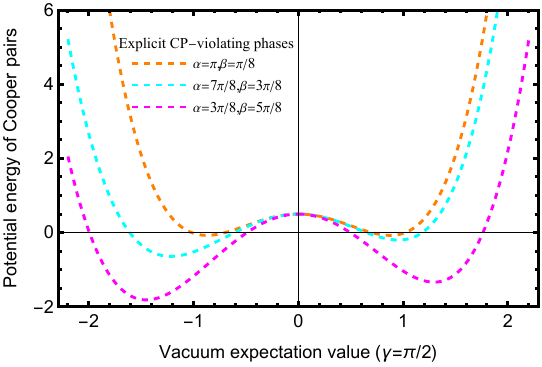}
\caption{Asymmetric double well}
\label{fig1b}
\end{subfigure} 
\caption{(color online) Potential energy of Cooper pairs in the representative case of explicit CP-violating phases.}
\label{fig1}
\end{figure*}

\begin{figure*}[t]
\centering
\begin{subfigure}[h]{0.49\textwidth}
\centering
\includegraphics[width=\textwidth]{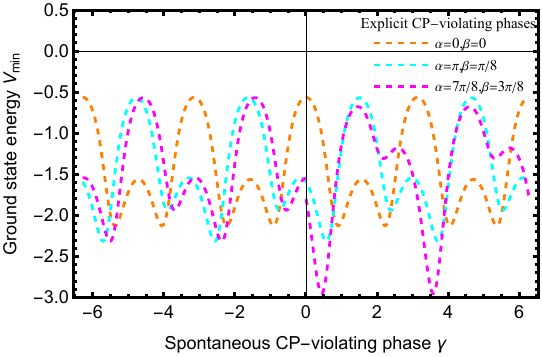}
\caption{Ground state energy}
\label{fig2a}
\end{subfigure}
\hfill
\centering
\begin{subfigure}[h]{0.49\textwidth}
\centering         
\includegraphics[width=\textwidth]{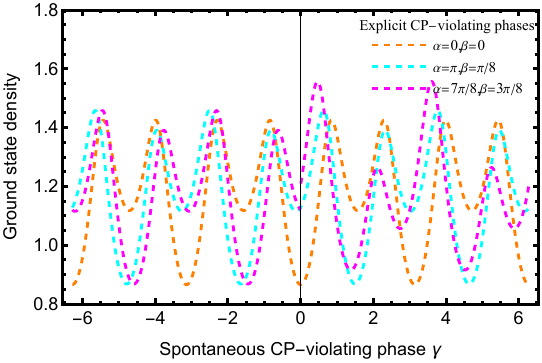}
\caption{Ground state density}
\label{fig2b}
\end{subfigure} 
\caption{(color online) Ground state energy and density of Cooper pairs in the representative case of explicit CP-violating phases.}
\label{fig2}
\end{figure*}

\subsection{Explicit CP-violation \label{subsec2}}

Next, we transfer our attention to the explicit CP-violation and its CP-violating signatures. We assume that all the CP-violating phases ($\alpha$, $\beta$, and $\gamma$) are present. The scalar potential evaluated at $\langle \phi \rangle \equiv e^{i \gamma}v/\sqrt{2}$ satisfies the following expression:
\begin{equation}
\begin{split}
V\big\vert_{\phi =\langle \phi \rangle} =& \frac{1}{2}\lambda_0 v^2 + \frac{1}{4}\lambda_1 v^4 +  \lambda_2v^2 \cos(\alpha + 2\gamma) \\
&+ \frac{1}{2} \lambda_3v^4 \cos(\beta + 4\gamma).    
\end{split}
\end{equation}
In order to reveal the non-trivial role of the three phases in CP-violation, the minimum of the scalar potential needs to be identified. The extrema of the scalar potential can be obtained by differentiating the above expression with respect to the VEV ($v$) and the phase ($\gamma$): 
\begin{align}
&\begin{aligned}
\frac{\partial V}{\partial v} \Bigg\vert_{\phi = \langle \phi \rangle} =& \lambda_0 v + \lambda_1 v^3 +  2\lambda_2v \cos(\alpha + 2\gamma) \\
&+ 2\lambda_3v^3 \cos(\beta + 4\gamma)=0,
\label{dvdv}
\end{aligned}\\
&\begin{aligned}
\frac{\partial V}{\partial \gamma}\Bigg\vert_{\phi = \langle \phi \rangle}=&\lambda_2 \sin(\alpha + 2\gamma)+ \lambda_3 v^2 \sin(\beta + 4\gamma) = 0. 
\label{dvdgamma}
\end{aligned}
\end{align}
The solutions to the above two equations do not necessarily correspond to the minimum points, but instead it may correspond to the stationary points, including the maximum, minimum, or saddle points. To be the minimum points, some additional conditions should be satisfied and such conditions would also impose further constraints on the parameter space.

Although, in principle, finding the full solutions to Eq. (\ref{dvdv}) and (\ref{dvdgamma}) is possible, the solutions are so cumbersome that it is difficult to examine and display them. To overcome the difficulty, some approximations of the solutions can be applied. For instance, Ref. \cite{mcdonald1995cosmological} shows that the analytical solutions can be presented under the assumption that $\alpha$ and the other parameter are small. However, Ref. \cite{haber2012group} shows that the phase $\alpha$ can take the value as large as $\pm \pi/4$. The analysis of the restrictions on the CP-violating phases does not necessarily depend on the exact solutions to Eq. (\ref{dvdv}) and (\ref{dvdgamma}). Instead, it is possible to obtain the restrictions on the CP-violating phases by examining the equations of the stationary points separately. For the purpose of this study, we prefer to focus on analyzing the trends of such equations rather than finding the exact solutions to them. Such an approach has been used to derive the conditions on spontaneous CP-violation (see. e.g. Ref. \cite{haber2012group}).

Although it is too cumbersome to display and examine the full solutions of Eq. (\ref{dvdv}) and (\ref{dvdgamma}), it is still possible to gain some insights into the origin and the signatures of CP-violation by analyzing the condition in Eq. (\ref{dvdv}) at a fixed value of $\gamma$. The solutions to Eq. (\ref{dvdv}) can be given by
\begin{equation}
v^2 = -\frac{\lambda_0 + 2\lambda_2 \cos{(\alpha + 2\gamma)}}{\lambda_1 + 2\lambda_3\cos{(\beta + 4\gamma)}}.
\label{vev}
\end{equation}
The minimum value of the scalar potential satisfies the following expression:
\begin{equation}
V_{\text{min}} = -\frac{\big[\lambda_0 + 2\lambda_2 \cos{(\alpha + 2\gamma)}\big]^2}{4\lambda_1 + 8\lambda_3\cos{(\beta + 4\gamma)}}.
\label{emin}
\end{equation}
Since the initial phase ($\gamma$) of the VEV would pick up a minus sign under the CP transformation \cite{cheung2012bubble}, we have indicated explicitly that the minimum value depends on the phase $\gamma$. 

\begin{figure*}[t]
\centering
\begin{subfigure}[h]{0.49\textwidth}
\centering
\includegraphics[width=\textwidth]{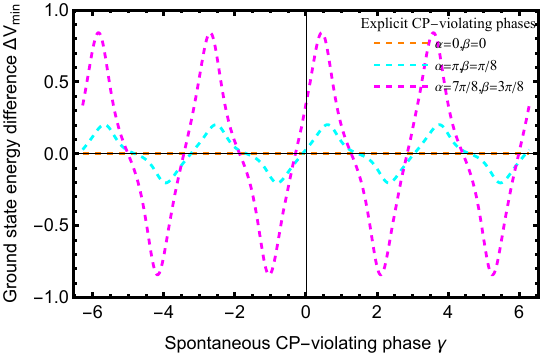}
\caption{Energy difference of ground sates}
\label{fig3a}
\end{subfigure}
\hfill
\centering
\begin{subfigure}[h]{0.49\textwidth}
\centering         
\includegraphics[width=\textwidth]{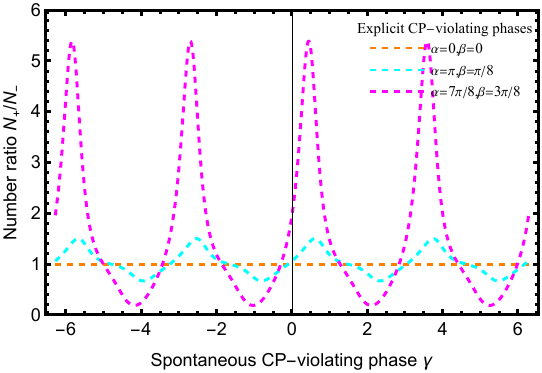}
\caption{Number ratio of ground sates}
\label{fig3b}
\end{subfigure} 
\caption{(color online) Energy difference and number ratio of Cooper pairs in the representative case of explicit CP-violating phases.}
\label{fig3}
\end{figure*}

The CP-transformed scalar potential evaluated at $\langle \phi \rangle^{\prime} \equiv e^{i (\delta-\gamma)}v/\sqrt{2}$ satisfies the expression\footnote{In addition to the spontaneous CPV phase factor $e^{-i\gamma}$, the antineutron Cooper field would gain an additional phase factor $e^{i\delta}$ due to the CP transformation.}:
\begin{equation}
\begin{split}
V^\text{CP}\big\vert_{\phi =\langle \phi \rangle^{\prime}} =& \frac{1}{2}\lambda_0 v^2 + \frac{1}{4}\lambda_1 v^4 +  \lambda_2v^2 \cos(\alpha - 2\gamma +2\delta) \\
&+ \frac{1}{2} \lambda_3v^4 \cos(\beta - 4\gamma + 4\delta)\\
=& \frac{1}{2}\lambda_0 v^2 + \frac{1}{4}\lambda_1 v^4 +  \lambda_2v^2 \cos(-\alpha - 2\gamma) \\
&+ \frac{1}{2} \lambda_3v^4 \cos(\beta -4\alpha - 4\gamma),
\end{split}
\end{equation}
or,
\begin{equation}
\begin{split}
V^\text{CP}\big\vert_{\phi =\langle \phi \rangle^{\prime}} =& \frac{1}{2}\lambda_0 v^2 + \frac{1}{4}\lambda_1 v^4 +  \lambda_2v^2 \cos(\alpha -\beta - 2\gamma) \\
&+ \frac{1}{2} \lambda_3v^4 \cos(-\beta - 4\gamma).
\end{split}
\end{equation}
Here, the auxiliary phase $\delta$ comes from the rephrasing transformation. We have made the term associated with $\lambda_2$ symmetric under the CP transformation by defining $\delta \equiv -\alpha$. Similarly, we can also make the term associated with $\lambda_3$ symmetric under the CP transformation by defining $\delta \equiv -\beta/2$. After the CP transformation, the minimum value evaluated at $\langle \phi \rangle^{\prime} \equiv e^{i (\delta-\gamma)}v/\sqrt{2}$ satisfies the expression: 
\begin{equation}
\begin{split}
V_{\text{min}}^{\text{CP}} &= -\frac{\big[\lambda_0 + 2\lambda_2 \cos{(-\alpha - 2\gamma)}\big]^2}{4\lambda_1 + 8\lambda_3\cos{(\beta - 4\gamma - 4\alpha)}},
\end{split}
\label{emincp}
\end{equation}
or,
\begin{equation}
\begin{split}
V_{\text{min}}^{\text{CP}} &= -\frac{\big[\lambda_0 + 2\lambda_2 \cos{(\alpha -\beta - 2\gamma)}\big]^2}{4\lambda_1 + 8\lambda_3\cos{(-\beta - 4\gamma)}}.
\end{split}
\label{emincp2}
\end{equation}
The potential energy of Cooper pairs in the representative case of explicit CP-violating phases in Fig. \ref{fig1}. The ground state energy and density of Cooper pairs in the representative case of explicit CP-violating phases in Fig. \ref{fig2}. The ratio between the number of neutron and antineutron bubbles can be given by \cite{comelli1994spontaneous}
\begin{equation}
\frac{N_{+}}{N_{-}} \equiv e^{\frac{\Delta V}{T}}.
\end{equation}
Here, the energy difference $\Delta V$ between the true and false vacua can be evaluated more conveniently using Eq. (\ref{emincp2}):
\begin{equation}
\begin{split}
\Delta V \equiv& V_{\text{min}}-V_{\text{min}}^{\text{CP}}\\
=& -\frac{\Big[2\lambda_0 + 8\lambda_2 \cos{\Big(\frac{2\alpha -\beta}{2}\Big)}\cos{\Big(\frac{\beta + 4\gamma}{2}\Big)}\Big]}{4\lambda_1 + 8\lambda_3\cos{(\beta + 4\gamma)}}\\
&\times\sin{\Big(\frac{2\alpha -\beta}{2}\Big)}\sin{\Big(\frac{\beta + 4\gamma}{2}\Big)}, 
\end{split}
\end{equation}
The energy difference and number ratio of Cooper pairs in the representative case of explicit CP-violating phases are presented in Fig. \ref{fig3}.

Before we proceed to explore the signatures of the explicit CP-violation, we will demonstrate its origin by examining whether all the CP-violating phases can be rotated away and whether the minimum of the scalar potential is symmetric with respect to the sign-flipping of $\gamma$. The conditions for CP-violation in the models with the scalar fields have been intensively studied for years (see e.g. Refs \cite{mcdonald1995cosmological,branco1999cp}). More recent and comprehensive analyses can be found in Refs. \cite{haber2012group,branco2016group}. Nevertheless, a brief review of the relevant results in the literature might be helpful for a better understanding of the CP-violating signatures. Again, we will pay more attention on the physical consequences of the CP violation, such as the $nn \rightarrow \bar{n}\bar{n}$ transition process. In case of $\lambda_2 = 0$ and $\lambda_3 = 0$, Eq. (\ref{eqvv}) becomes the usual scalar potential in the GL model. Since the CP-violating phases do not appear in the scalar potential, we could obtain $V_{\text{min}}^{\text{CP}}=V_{\text{min}}$ and no CP-violation could occur. The manifold of the vacuum contains two disconnected points: $v_1 = \pm \sqrt{-\lambda_0/\lambda_1}$. In the absence of CP-violation, since it is allowed to perform a field redefining transformation (i.e. $\phi \mapsto -\phi$) without causing any observable physical consequences, we could only choose the real and positive value of the VEV \cite{coito2021dark}. This is a trivial case which we are not interested in. We organize our discussion in the following three non-trivial cases: (1) $\lambda_2 = 0$, $\lambda_3 \neq 0$; (2) $\lambda_2 \neq 0$, $\lambda_3 = 0$; (3) $\lambda_2 \neq 0$, $\lambda_3 \neq 0$.

In case (1), the CP-violating phases ($\alpha$ and $\gamma$) appear in the scalar potential. It is always possible to remove the CP-violating phases ($\alpha$ and $\gamma$) away by adjusting the phase $\delta$ properly. Furthermore, it is sufficient to make the scalar potential symmetric with respect to the sign-flipping of $\gamma$ ($\gamma \mapsto -\gamma$) by defining $\delta \equiv -\alpha$. Therefore, it is possible to obtain $V_{\text{min}}^{\text{CP}}=V_{\text{min}}$ so that CP-violation could not occur. Up to an irrelevant phase, the vacuum manifold contains two disconnected points: $v_1 = \pm \sqrt{-(\lambda_0+2\lambda_2)/\lambda_1}$.

In case (2), the CP-violating phases ($\beta$ and $\gamma$) appear in the scalar potential. It is always possible to remove $\beta$ and $\gamma$ away by adjusting the phase $\delta$ properly. Furthermore, it is also sufficient to make the scalar potential symmetric under the transformation that flips the sign of the phase (i.e. $\gamma \mapsto -\gamma$) by defining $\delta \equiv -\beta/2$. Therefore, it is possible to obtain $V_{\text{min}}^{\text{CP}}=V_{\text{min}}$ so that CP-violation could not occur. Up to an irrelevant phase, the vacuum manifold contains two disconnected points: $v_2 = \pm \sqrt{-\lambda_0)/(\lambda_1+2\lambda_3)}$. In this case, the scalar potential has an additional $Z_4$ symmetry under the transformation: $\phi \mapsto e^{i2\pi k/4}\phi$ ($k=0$, $1$, $2$, $3$).

In the above two cases, the vacuum phase $\gamma$ is not an observable variable. The vacua ($\pm v_{i}e^{i\gamma}$, $i=1$, $2$) share degenerate energy levels with the vacua that differ by a sign-flipping of $\gamma$ ($\pm v_{i}e^{-i\gamma}$, $i=1$, $2$). This suggests that all the vacua can be considered as originated from the real component of $\phi$ [i.e. Re($\phi$)]. Since each pair of vacua exhibits a two-fold degeneracy, the $\mathcal{B}$-asymmetry induced by the tunneling between these two pairs of vacua tends to cancel out so that the $nn \rightarrow \bar{n}\bar{n}$ transition process would not occur. In this case, the bubbles of the neutron superfluid have the same population with that of the antineutron superfluid.

In case (3), in order to obtain a CP-invariant result, the phase $\delta$ has to satisfy two conditions: $\delta \equiv -\alpha$ and $\delta \equiv -\beta/2$. For a generic phase $\delta$, it would not be possible to satisfy these two conditions simultaneously unless the phases $\alpha$ and $\beta$ satisfy the relationship: $2\alpha -\beta = 2k\pi$ ($k \in \mathbb{N}$). Therefore, if such a relationship is not satisfied, the additional phases associated with CP-violation cannot be removed way by a rephrasing transformation. In this case, as it has been shown in Ref. \cite{haber2012group}, it is possible to obtain $V_{\text{min}}^{\text{CP}}\neq V_{\text{min}}$ so that CP-violation can be established. The vacuum that differs by a sign-flipping of $\gamma$ would have different energies and thus the two-fold degeneracy of each vacuum could be removed. The tunneling from the neutron superfluid to the antineutron superfluid cannot cancel out the tunneling in the opposite direction, favoring the production of neutron Cooper pairs over the production of antineutron Cooper pairs, or vice versa. Therefore, the $nn \rightarrow \bar{n}\bar{n}$ transition process would occur.

The $nn \rightarrow \bar{n}\bar{n}$ transition process would only occur if both of the $\mathcal{B}$-violating terms in Eq. (\ref{eqvv}) are present. However, the $\mathcal{B}$-violating terms would not necessarily lead to CP-violation unless some conditions are satisfied. If one of the two $\mathcal{B}$-violating terms in Eq. (\ref{eqvv}) is absent or if the condition $2\alpha -\beta = 2k\pi$ is satisfied, then it is always possible to remove the CP-violating phases by a rephrasing transformation so that no CP violation would occur. A more comprehensive analysis of the conditions for CP-violation on the number of the broken $U(1)$ phases and the number of complex coupling parameters can be found in Refs. \cite{haber2012group,branco2016group}. In the following discussion, unless otherwise specified, we will assume that $2\alpha -\beta = 2k\pi$ is not satisfied and only consider the CP-violating case.

In what follows, we will demonstrate the physical consequences arising from the CP-violating phases. At the beginning of this subsection, we have illustrated that the interplay of the three phases (i.e. $\alpha$, $\beta$, and $\gamma$) could lead to the explicit CP-violation. One of the most direct physical consequences of the CP-violation is the asymmetry between neutron superfluids and antineutron superfluids in the physical quantities, such as VEV, mass, and etc. Since the VEV is associated with the number density ($N_c$) of the Cooper pairs \cite{annett2004superconductivity,tinkham1996introduction} and is closely related to the effective mass, we will explore the properties of the VEV first.

As can be seen in Eq. (\ref{vev}), the VEV depends on the CP-violating phases explicitly. Following Refs. \cite{haber2012group,branco2016group}, unless the condition $2\alpha -\beta = 2k\pi$ is satisfied, it would not be possible to remove all the three phases simultaneously by a rephrasing transformation. Up to an irrelevant phase, the neutron superfluid can be related to the antineutron superfluid by a CP (or a charge-conjugation) transformation. The spontaneous CPV phase for the antineutron Cooper field would differ from the one for the neutron Cooper field by a minus sign. Therefore, the VEV of the neutron superfluid would be different from the VEV of the antineutron superfluid: $v_3 \neq v_3^{\text{CP}}$. This illustrates that not only the infinite degeneracy associated with the $U(1)$ symmetry is removed, but also the double degeneracy associated with the sign-flipping of $\gamma$ is removed. Furthermore, since the VEV is associated with the number density ($N_c$) of the Cooper pairs \cite{annett2004superconductivity,tinkham1996introduction}, Eq. (\ref{vev}) suggests that, for the ground state, the number density of the neutron Cooper pairs would be different from that of the antineutron Cooper pairs. Such effects can be considered as a manifestation of CP-violation and may lead to observable effects.

The mass squared matrix for the real and imaginary components can be given by
\begin{equation}
M^2 = 
\left(
\begin{array}{cc}
M^2_{11} & M^2_{12}\\
M^2_{21} & M^2_{22}
\end{array}
\right),
\label{msq}
\end{equation}
where the matrix has a transpose symmetry: $ M^2_{12}=M^2_{21}$. The matrix elements can be obtained at the tree level by taking the second derivative of the scalar potential with respect to the scalar field:
\begin{equation}
M^2_{ij} \equiv \frac{\partial^2V}{\partial\phi_i\partial\phi_j} \Bigg\vert_{\phi = \langle \phi \rangle} \equiv 0, \quad i=1,2, 
\end{equation}
with
\begin{align}
&\begin{aligned}
M_{11}^2 =& \lambda_0 + 2\lambda_1 v^2 + 2 \lambda_2\cos{\alpha} + \lambda_1 v^2 \cos{(2\gamma)}\\
&+6\lambda_3 v^2 \cos{(\beta + 2\gamma)},
\label{M11sq}
\end{aligned}\\
&\begin{aligned}
M_{12} =& \lambda_1 v^2 \sin{(2\gamma)} -2 \lambda_2\sin{\alpha} -6\lambda_3 v^2 \sin{(\beta + 2\gamma)},
\label{M12sq}
\end{aligned}\\
&\begin{aligned}
M_{22}^2 = &\lambda_0 + 2\lambda_1 v^2 - 2 \lambda_2\cos{\alpha} - \lambda_1 v^2 \cos{(2\gamma)}\\
&-6\lambda_3 v^2 \cos{(\beta + 2\gamma)},
\label{eqm2}
\end{aligned}
\end{align}

As can be seen, the real and imaginary components couple to each other and thus are not mass eigenstates. The mass eigenstates can be obtained by an orthogonal transformation $T$,
\begin{equation}
\begin{split}
\left(
\begin{array}{c}
\Phi_1\\
\Phi_2
\end{array}
\right)
&= T(\theta)
\left(
\begin{array}{c}
\phi_1\\
\phi_2
\end{array}
\right)\\
&=\left(
\begin{array}{cc}
\cos{\theta} & \sin{\theta} \\
-\sin{\theta} & \cos{\theta}
\end{array}
\right)
\left(
\begin{array}{c}
\phi_1\\
\phi_2
\end{array}
\right).
\end{split}
\end{equation}
Here, $\Phi_1$ and $\Phi_2$ are the mass eigenstates. Accordingly, the mass squared matrix can be formally diagonalized as 
\begin{equation}
\begin{split}
&T(\theta) M^2 T(\theta)^{-1} \\
&= 
\left(
\begin{array}{cc}
\cos{\theta} & \sin{\theta} \\
-\sin{\theta} & \cos{\theta}
\end{array}
\right)
\left(
\begin{array}{cc}
M_{11}^2 & M_{12}^2\\
M_{12}^2 & M_{22}^2
\end{array}
\right)
\left(
\begin{array}{cc}
\cos{\theta} & \sin{\theta} \\
-\sin{\theta} & \cos{\theta}
\end{array}
\right)^{-1}\\
&=
\left(
\begin{array}{cc}
m_1^2(\alpha, \beta, \gamma) & 0\\
0 & m_2^2(\alpha, \beta, \gamma)
\end{array}
\right).    
\end{split}
\label{tm2t}
\end{equation}
Here, for the convenience of analyzing the CP-violating signatures, we have indicated that the diagonal matrix elements depend on the three phases explicitly. The mixing angle $\theta$ between the real and imaginary components can be defined by 
\begin{equation}
\tan{2\theta} \equiv \frac{2M_{12}^2}{M_{11}^2 - M_{22}^2}.
\label{tant}
\end{equation}
The masses for the mass eigenstates can be given by

\begin{equation}
 m_{1,2}^2 =\frac{1}{2} \Big[M_{11}^2 + M_{22}^2 \pm \sqrt{(M_{11}^2 - M_{22}^2)^2 + 4(M_{12}^2)^2}   \Big]. 
 \label{m12dig}
\end{equation}
In Eq. (\ref{eqvv}), in the absence of the $\mathcal{B}$-violating terms, the system possess a continuous $U(1)$ symmetry. However, in the presence of the $\mathcal{B}$-violating terms, the system does not possess the continuous $U(1)$ symmetry but instead it respects a discrete $Z_2$ symmetry. According to the Nambu-Goldstone (NG) theorem \cite{nambu1961dynamical,goldstone1961field,goldstone1962broken}, the spontaneous breaking of a continuous symmetry gives rise to massless NG bosons. In the scenario with discrete symmetries, the spontaneous breaking of a discrete symmetry gives rise to massive Pseudo-Nambu-Goldstone (PNG) bosons \cite{haber2012group}. As can be seen from Eq. (\ref{m12dig}), both $m_1^2$ and $m_2^2$ are non-zero. If the coupling parameters of the $\mathcal{B}$-violating terms are small (i.e. $\lambda_{2,3} \ll \lambda_{0,1}$), we would expect $m_1^2 >> m_2^2$, where $m_1^2$ is associated with the effective mass of the neutron Cooper field and $m_2^2$ is associated with the effective mass of the PNG boson. The PNG boson gets a quite small but non-zero mass due to the spontaneous breaking of the $Z_2$ symmetry.

\begin{table*}[t]
\caption{The possible lowest energies of the vacuum (the ground state) in various cases under the conditions: $\lambda_{0} < 0$, $\lambda_{1} > 0$, and $\lambda_{0,1} \gg \lambda_{2,3}$ ($k \in \mathbb{N}$).
}
\begin{ruledtabular}
\begin{tabular}{l|l|l|l|l}
\diagbox{Conditions}{Cases} & $\lambda_{2} > 0$, $\lambda_{3} > 0$ & $\lambda_{2} > 0$, $\lambda_{3} < 0$ & $\lambda_{2} < 0$, $\lambda_{3} > 0$ &$\lambda_{2} < 0$, $\lambda_{3} < 0$\\\hline
$\alpha$, $\gamma$&$\alpha+2\gamma=(2k+1)\pi$&$\alpha+2\gamma=(2k+1)\pi$&$\alpha+2\gamma=2k\pi$&$\alpha+2\gamma=2k\pi$\\\hline
$\beta$, $\gamma$ &$\beta+4\gamma=(2k+1)\pi$&$\beta+4\gamma=2k\pi$&$\beta+4\gamma=(2k+1)\pi$&$\beta+4\gamma=2k\pi$\\\hline
$V_{\text{min}}$&$-\frac{(\lambda_0 - 2\lambda_2)^2}{4\lambda_1 - 8\lambda_3}$&$-\frac{(\lambda_0 - 2\lambda_2)^2}{4\lambda_1 + 8\lambda_3}$&$-\frac{(\lambda_0 + 2\lambda_2)^2}{4\lambda_1 - 8\lambda_3}$&$-\frac{(\lambda_0 + 2\lambda_2)^2}{4\lambda_1 + 8\lambda_3}$\\
\end{tabular}
\end{ruledtabular}
\label{tab2}
\end{table*}

In the above analysis, we have shown that it is not possible to completely rotate the CP-violating phases away by a field redefinition. Up to an irrelevant phase, the antineutron Cooper field is different from the neutron Cooper field by a sign-flipping of $\gamma$. We denote the effective mass of the antineutron Cooper field by $m_1^{2\text{CP}} \equiv m_1^2(\alpha, \beta, \gamma^{\prime})$ ($\gamma^{\prime}\equiv \delta-\gamma$). The effective mass of the antineutron Cooper field would be different from that of the neutron Cooper field: $ m_1^{2\text{CP}} \neq m_1^2$. As can be seen, the mass asymmetry between the neutron Cooper field and the antineutron Cooper field is not only originated from the CP-violating phases but also originated from the mixing between the real and imaginary components. Furthermore, except for $M_{11}^2$, all the matrix elements in Eq. (\ref{msq}) are very small. Both the mixing angle and the CP-violating effect resulted from the mixing are very small.

From the experimental aspects of particle physics, the CP properties of the Higgs-like boson have not been completely pinned down so far \cite{aad2020cp,sirunyan2020measurements}. Although a pure CP-odd hypothesis has been excluded, the Higgs-like boson as an admixture of CP-even and CP-odd components could still be possible \cite{aad2020cp,sirunyan2020measurements}. Since the phenomena in condensed matter physics has a much clearer interpretation and a more intuitive picture, the analogies between these two fields may help explain some intrinsic puzzles and even clarify some fundamental issues in particle physics. Although some phenomena that have been observed in condensed matter physics may not have their counterparts in particle physics (or vice versa), such analogies could provide helpful clues and insights into future investigations of the BEH model.

\section{False vacuum decay \label{sec4}}

Before proceeding further, we identify the lowest possible energies of the vacuum state (or the ground state) of the system, which satisfy the expression given by Eq. (\ref{emin}). Again, as can be seen, the energy of the vacuum state also depends on the three phases, which cannot be removed away simultaneously unless the condition $2\alpha -\beta = 2k\pi$ is satisfied. Since the neutron superfluid is related to the antineutron superfluid by a CP transformation, where the latter acquires an extra phase factor $e^{i\delta}$ besides the sign-flipping of $\gamma$. Furthermore, we have illustrated in the previous discussion that such CP-violating phases cannot be completely rotated away. Apparently, Eq. (\ref{emin}) shows that $V_{\text{min}}^{\text{CP}} \neq V_{\text{min}} $. This suggests that the vacua of the neutron superfluid and the antineutron superfluid have different energies.

In quantum field theory, a vacuum can be classified as either a true vacuum or a false vacuum, depending on their energies. A true vacuum has a relatively lower energy, comparing with a false vacuum. Similarly, if the vacuum of the neutron superfluid has a lower energy than the vacuum of the antineutron superfluid, it can be classified as a true vacuum otherwise it can be classified as a false vacuum. Tab. \ref{tab2} shows the possible lowest energies of the vacuum state and the conditions under which such the lowest energies can be obtained. Such conditions show explicitly that if the CP-violating phases cannot be rotated away by a phase redefinition, the degeneracy of the vacuum states related by a sign-flipping of $\gamma$ would be removed. If the vacuum of the neutron satisfies the conditions presented in Tab. \ref{tab2}, it can be classified as the true vacuum. In this case, the vacuum of the antineutron superfluid can be classified as the false vacuum because it would not satisfy these conditions and thus would have a relatively higher energy. The false vacuum would be unstable and tunnel into the true vacuum. Therefore, the $nn \rightarrow \bar{n}\bar{n}$ transition process can be manifested as a false vacuum decay through a quantum tunneling process induced by the superfluid pairing interactions. If the vacuum of the neutron superfluid is energetically more favorable and the bubbles or domains containing the neutron superfluid would expand and push the domain walls outward, while the domains containing antineutron superfluid would shrike and the corresponding domain walls would collapse \cite{dvali1995biased}.

The vacuum in quantum field theory has a very complicated structure and a remarkably rich physical content. In quantum field theory, vacuum is defined as the state with the lowest possible energy. The false vacuum decay refers to the decay of a metastable ground state into a stable ground state through e.g. the nucleation of spatially localized bubbles or domains (see e.g. Ref. \cite{gleiser2008bubbling}). The calculation of the false vacuum decay rate has been pioneeringly developed in Refs. \cite{coleman1977fate,*coleman1977erratum,callan1977fate,kobsarev1974bubbles}. In a gauge theory, the calculation of the false vacuum decay rate can be found in e.g. Refs. \cite{endo2017false,chigusa2020precise}. A more comprehensive treatment of the false vacuum decay rate can be found in Ref. \cite{andreassen2017precision}. An introductory review of the false vacuum decay can be found in Ref. \cite{devoto2022false}. The false vacuum decay rate per unit volume can be expressed by \cite{coleman1977fate,callan1977fate,coleman1980gravitational,andreassen2016direct,andreassen2017precision}
\begin{equation}
 \frac{\Gamma}{V} \simeq A e^{-B}.
\end{equation}
The decay rate was first derived in Refs. \cite{coleman1977fate,callan1977fate,coleman1980gravitational} and recently rederived using an alternate approach in Refs. \cite{andreassen2016direct,andreassen2017precision}.

Spontaneous symmetry breaking is one of the key ideas that carry rich implications in both the particle physics and condensed matter physics. The system has a family of vacua that can be related to each other by the transformation of the symmetry group. Usually, the true ground state (or vacuum) is not any one among these states. Instead, it may be a superposition of these states if the amplitude for the transition between these states is non-vanishing, due to the tunneling process through potential barrier \cite{maggiore2005modern}. The tunneling probability is exponentially suppressed by the spatial volume of the system. In particle physics. the tunneling process occurs in the space of infinite volume and thus the probability of the tunneling between two vacua is vanishing and there is no mixing between the two vacua. Although the analogies between particle physics and condensed matter physics are commonly used, some subtle differences deserve special attention. In the space of infinite volume, the matrix elements between different vacuum states tend to be vanishing \cite{weinberg1996quantum,maggiore2005modern}. The vacuum associated with the neutron superfluid is separated from the vacuum associated with the antineutron superfluid by a potential barrier. Analogies to the case in quantum field theory, there is a a tunneling process between the two vacua.
However, the situation is significantly different in condensed matter physics as the system under discussion usually has a finite spatial size or volume, making the tunneling process between the two vacua possible. For example, a superfluid system in a neutron star should have a size smaller than the volume of the neutron star. The $nn \rightarrow \bar{n}\bar{n}$ transition process as a manifestation of CP-violation would occur inside neutron stars. the From experimental aspects, a possible signal for the false vacuum decay via bubble formation in ferromagnetic superfluids has been observed \cite{berti2023observation}. It is natural to anticipate that the signals for the $nn \rightarrow \bar{n}\bar{n}$ transition process can be possibly detected through an astrophysical observation.

After the discovery of the Higgs-like boson \cite{aad2012observation,chatrchyan2012observation,chatrchyan2013observation}, attentions are brought back to the historical connections and formal analogies among the BEH model \cite{englert1964broken,higgs1964broken1,higgs1964broken2,guralnik1964global}, GL model \cite{ginzburg1950on,ginzburg2009superconductivity}) and the BCS models \cite{bardeen1957theory}. The first one plays role mainly in the field of particle physics whereas the other two play role mainly in the field of condensed matter physics. Both fields have surprisingly similar parallel counterparts in their physical concepts and mathematical structure. Since the phenomena in condensed matter physics has a much clearer interpretation and a more intuitive picture, such analogies may help explain some intrinsic puzzles and even clarify some fundamental issues in particle physics. The analogies between particle physics and condensed matter physics have yielded fruitful results in both fields. In particular, some special analogies have applications in new-physics searches. For example, they can be useful for predicting the mass of extra Higgs bosons \cite{volovik2014higgs}. From the theoretical aspects, the exact shape of the Higgs is still unknown even at the tree level \cite{agrawal2020determining}. Although some phenomena that have been observed in condensed matter physics may not have their counterparts in particle physics (or vice versa), such analogies could provide helpful insights or clues for future investigations of the BEH model.

\section{Summary and Outlook}

In this work, we have reviewed the historical connections and formal analogies among the BEH model \cite{englert1964broken,higgs1964broken1,higgs1964broken2,guralnik1964global}, GL model \cite{ginzburg1950on,ginzburg2009superconductivity}) and the BCS models \cite{bardeen1957theory}. By invoking such analogies, we have explored a new picture of $\mathcal{B}$-violation. A possible manifestation of this new picture includes the transition between a pair of neutrons and a pair of antineutrons (i.e. $nn \rightarrow \bar{n}\bar{n}$), which violates baryon number by 4 units (i.e. $|\Delta\mathcal{B}|=4$). To be more specific, the $nn \rightarrow \bar{n}\bar{n}$ process can be considered as a false vacuum decay through a quantum tunneling process induced by the superfluid pairing interactions. We have argued that both the $\mathcal{B}$-violating and CP-violating effects can be quite naturally accommodated in the $nn \rightarrow \bar{n}\bar{n}$ process.

There are still many mysteries of neutron that need to be revealed. The $n$-$\bar{n}$ oscillation, which violates baryon number by 2 units (i.e. $|\Delta\mathcal{B}|=2$), can be mediated by new heavy bosons or can be resulted from high-dimensional operators. Therefore, this process is greatly suppressed by the energy scale of new physics. The direct searches for new heavy bosons at the LHC have yielded trivial results, suggesting that the energy scale of new physics tends to be so high that a direct laboratory detection might be inappropriate through the current experimental techniques. By contrast, neutron stars are one of the densest objects in our universe and contain so many neutrons that a tiny $\mathcal{B}$-violating effect can be amplified, making it possible to search for such signals through astrophysical observations.

At high temperature, the neutrons in neutron stars are fluctuating and disorganized. In this case, there would be no preferred direction and the system possesses a $SO(3)$ symmetry. When the temperature of neutron stars decreases below a critical temperature $T_c$, a superfluid phase transition may occur and neutrons tend to form Cooper pairs. The spins of the paired neutrons in a s-wave spin-singlet state point to the opposite directions and tend to cancel out each other. In this case, the s-wave spin-singlet neutron Cooper pair can be modeled by a complex scalar field called the neutron Cooper field, which carries two units of baryon number. After the phase transition, the system can be described by the GL model, which possess a $U(1)$ symmetry. As mentioned earlier, the AD field share some similarities with the neutron Cooper field. The AD mechanism may provide some useful insights or clues on the form of the scalar potential characterized by $\mathcal{B}$-violation for the neutron superfluids.

In the models with new scalars, CP-violation can be mainly achieved either explicitly through the complex coupling parameters (or the complex phase factors of the coupling parameters) in the Lagrangian (density) and/or through complex vacuum expectation values (or the complex phase factors of vacuum expectation values) after the spontaneous symmetry breaking \cite{sozzi2007discrete,bigi2009cp}. In the presence of the $\mathcal{B}$-violating terms, the neutron superfluid does not possess a continuous $U(1)$ symmetry but instead it respect a discrete $Z_2$ symmetry. We have reaffirmed the claim that the CP-violating effect can only occur if both of the $\mathcal{B}$-violating terms are present. In this case, we have incorporated three CP-violating phases into the model such as the two explicit CPV phases (i.e. $\alpha$ and $\beta$) from the complex coupling parameters and one spontaneous CPV phase (i.e. $\gamma$) from the complex vacuum expectation value (VEV). We have illustrated that the interplay of the three phases (i.e. $\alpha$, $\beta$, and $\gamma$) leads to CP-violation. Specifically, such CP-violating phases cannot be removed way by a rephrasing transformation unless some conditions are satisfied.

One of the most direct physical consequences of the CP-violation is the asymmetry between the neutron superfluids and the antineutron superfluids in the physical quantities of interest, such as the VEV, mass, and etc. We have pointed out that the VEV of the antineutron superfluid can be different from the VEV of the neutron superfluid: $v^{\text{CP}} \neq v$, suggesting that not only the infinite degeneracy of the ground state arising from the $U(1)$ symmetry can be removed, but also the double degeneracy arising from the sign-flipping of $\gamma$ can be removed. Furthermore, since the VEV is associated with the number density ($N_c$) of the Cooper pairs \cite{annett2004superconductivity,tinkham1996introduction}, it also suggests that the number density of the neutron Cooper pairs would be different from that of the antineutron Cooper pairs.

Furthermore, we have indicated explicitly that the masses of the mass eigenstates are the functions of the three CPV phases. We have shown that it is not possible to completely rotate the CP-violating phases away by a field redefinition unless some conditions are satisfied. We have also demonstrated that there is an asymmetry between the effective mass of the antineutron Cooper field and that of the neutron Cooper field: $m_1^{2\text{CP}} \neq m_1^2$. The mass eigenstate is an admixture between the real and imaginary components of the neutron Cooper field. Furthermore, we have also shown that the PNG boson arising from the spontaneous breaking of a discrete symmetry gets a quite small but non-zero mass due to the spontaneous breaking of the $Z_2$ symmetry.

Finally, we have identified the lowest possible energies of the vacuum (or the ground state) of the system. Again, the energy of the vacuum also depends on the three CPV phases, which cannot be removed away simultaneously unless some conditions are satisfied. We have shown that the vacua of the antineutron superfluid and the neutron superfluid have different energies: $V_{\text{min}}^{\text{CP}} \neq V_{\text{min}}$. If the CP-violating phases cannot be rotated away by a field redefinition, the degeneracy of the vacuum states related by a sign-flipping of $\gamma$ would be removed. 
If the vacuum energy of the antineutron Cooper field is lower energy than that of the neutron Cooper field, then it would indicate that the vacuum of the antineutron superfluid is a true vacuum while the vacuum of the neutron superfluid is a false vacuum (or vice versa). Therefore, the $nn \rightarrow \bar{n}\bar{n}$ process can be manifested as a false vacuum decay through a quantum tunneling process induced by the superfluid pairing interactions. The transition rate for the $nn \rightarrow \bar{n}\bar{n}$ process can be estimated by the false vacuum decay rate, which can be modeled by an exponential function. The $\mathcal{B}$-violating process accompanied by CP-violation would open a promising avenue for exploring new physics effects beyond the SM.

\section*{Acknowledgement}

This work is supported by the National Natural Science Foundation of China (Grant No. 12104187 and 12022517), the Science and Technology Development Fund, Macau SAR (File No. 0048/2020/A1). The work of Y.L. Hao is supported by Macao Young Scholars Program (No. AM2021001), Jiangsu Provincial Double-Innovation Doctor Program (Grant No. JSSCBS20210940), and the Startup Funding of Jiangsu University (No. 4111710002). Y.L. Hao thanks Prof. Haijun Wang for many useful conversations.

\bibliography{reference}

\begin{thebibliography}{103}%
\makeatletter
\providecommand \@ifxundefined [1]{%
 \@ifx{#1\undefined}
}%
\providecommand \@ifnum [1]{%
 \ifnum #1\expandafter \@firstoftwo
 \else \expandafter \@secondoftwo
 \fi
}%
\providecommand \@ifx [1]{%
 \ifx #1\expandafter \@firstoftwo
 \else \expandafter \@secondoftwo
 \fi
}%
\providecommand \natexlab [1]{#1}%
\providecommand \enquote  [1]{``#1''}%
\providecommand \bibnamefont  [1]{#1}%
\providecommand \bibfnamefont [1]{#1}%
\providecommand \citenamefont [1]{#1}%
\providecommand \href@noop [0]{\@secondoftwo}%
\providecommand \href [0]{\begingroup \@sanitize@url \@href}%
\providecommand \@href[1]{\@@startlink{#1}\@@href}%
\providecommand \@@href[1]{\endgroup#1\@@endlink}%
\providecommand \@sanitize@url [0]{\catcode `\\12\catcode `\$12\catcode
  `\&12\catcode `\#12\catcode `\^12\catcode `\_12\catcode `\%12\relax}%
\providecommand \@@startlink[1]{}%
\providecommand \@@endlink[0]{}%
\providecommand \url  [0]{\begingroup\@sanitize@url \@url }%
\providecommand \@url [1]{\endgroup\@href {#1}{\urlprefix }}%
\providecommand \urlprefix  [0]{URL }%
\providecommand \Eprint [0]{\href }%
\providecommand \doibase [0]{http://dx.doi.org/}%
\providecommand \selectlanguage [0]{\@gobble}%
\providecommand \bibinfo  [0]{\@secondoftwo}%
\providecommand \bibfield  [0]{\@secondoftwo}%
\providecommand \translation [1]{[#1]}%
\providecommand \BibitemOpen [0]{}%
\providecommand \bibitemStop [0]{}%
\providecommand \bibitemNoStop [0]{.\EOS\space}%
\providecommand \EOS [0]{\spacefactor3000\relax}%
\providecommand \BibitemShut  [1]{\csname bibitem#1\endcsname}%
\let\auto@bib@innerbib\@empty
\bibitem [{\citenamefont {Aad}\ \emph {et~al.}(2012)\citenamefont {Aad},
  \citenamefont {Abajyan}, \citenamefont {Abbott}, \citenamefont {Abdallah},
  \citenamefont {Khalek}, \citenamefont {Abdelalim}, \citenamefont {Abdinov},
  \citenamefont {Aben}, \citenamefont {Abi}, \citenamefont {Abolins} \emph
  {et~al.}}]{aad2012observation}%
  \BibitemOpen
  \bibfield  {author} {\bibinfo {author} {\bibfnamefont {G.}~\bibnamefont
  {Aad}}, \bibinfo {author} {\bibfnamefont {T.}~\bibnamefont {Abajyan}},
  \bibinfo {author} {\bibfnamefont {B.}~\bibnamefont {Abbott}}, \bibinfo
  {author} {\bibfnamefont {J.}~\bibnamefont {Abdallah}}, \bibinfo {author}
  {\bibfnamefont {S.~A.}\ \bibnamefont {Khalek}}, \bibinfo {author}
  {\bibfnamefont {A.~A.}\ \bibnamefont {Abdelalim}}, \bibinfo {author}
  {\bibfnamefont {O.}~\bibnamefont {Abdinov}}, \bibinfo {author} {\bibfnamefont
  {R.}~\bibnamefont {Aben}}, \bibinfo {author} {\bibfnamefont {B.}~\bibnamefont
  {Abi}}, \bibinfo {author} {\bibfnamefont {M.}~\bibnamefont {Abolins}},  \emph
  {et~al.},\ }\href {\doibase 10.1016/j.physletb.2012.08.020} {\bibfield
  {journal} {\bibinfo  {journal} {Phys. Lett. B}\ }\textbf {\bibinfo {volume}
  {716}},\ \bibinfo {pages} {1} (\bibinfo {year} {2012})}\BibitemShut {NoStop}%
\bibitem [{\citenamefont {Chatrchyan}\ \emph {et~al.}(2012)\citenamefont
  {Chatrchyan}, \citenamefont {Khachatryan}, \citenamefont {Sirunyan},
  \citenamefont {Tumasyan}, \citenamefont {Adam}, \citenamefont {Aguilo},
  \citenamefont {Bergauer}, \citenamefont {Dragicevic}, \citenamefont
  {Er{\"o}}, \citenamefont {Fabjan} \emph
  {et~al.}}]{chatrchyan2012observation}%
  \BibitemOpen
  \bibfield  {author} {\bibinfo {author} {\bibfnamefont {S.}~\bibnamefont
  {Chatrchyan}}, \bibinfo {author} {\bibfnamefont {V.}~\bibnamefont
  {Khachatryan}}, \bibinfo {author} {\bibfnamefont {A.~M.}\ \bibnamefont
  {Sirunyan}}, \bibinfo {author} {\bibfnamefont {A.}~\bibnamefont {Tumasyan}},
  \bibinfo {author} {\bibfnamefont {W.}~\bibnamefont {Adam}}, \bibinfo {author}
  {\bibfnamefont {E.}~\bibnamefont {Aguilo}}, \bibinfo {author} {\bibfnamefont
  {T.}~\bibnamefont {Bergauer}}, \bibinfo {author} {\bibfnamefont
  {M.}~\bibnamefont {Dragicevic}}, \bibinfo {author} {\bibfnamefont
  {J.}~\bibnamefont {Er{\"o}}}, \bibinfo {author} {\bibfnamefont
  {C.}~\bibnamefont {Fabjan}},  \emph {et~al.},\ }\href {\doibase
  10.1016/j.physletb.2012.08.021} {\bibfield  {journal} {\bibinfo  {journal}
  {Phys. Lett. B}\ }\textbf {\bibinfo {volume} {716}},\ \bibinfo {pages} {30}
  (\bibinfo {year} {2012})}\BibitemShut {NoStop}%
\bibitem [{\citenamefont {Chatrchyan}\ \emph {et~al.}(2013)\citenamefont
  {Chatrchyan}, \citenamefont {Khachatryan}, \citenamefont {Sirunyan},
  \citenamefont {Tumasyan}, \citenamefont {Adam}, \citenamefont {Bergauer},
  \citenamefont {Dragicevic}, \citenamefont {Er{\"o}}, \citenamefont {Fabjan},
  \citenamefont {Friedl} \emph {et~al.}}]{chatrchyan2013observation}%
  \BibitemOpen
  \bibfield  {author} {\bibinfo {author} {\bibfnamefont {S.}~\bibnamefont
  {Chatrchyan}}, \bibinfo {author} {\bibfnamefont {V.}~\bibnamefont
  {Khachatryan}}, \bibinfo {author} {\bibfnamefont {A.~M.}\ \bibnamefont
  {Sirunyan}}, \bibinfo {author} {\bibfnamefont {A.}~\bibnamefont {Tumasyan}},
  \bibinfo {author} {\bibfnamefont {W.}~\bibnamefont {Adam}}, \bibinfo {author}
  {\bibfnamefont {T.}~\bibnamefont {Bergauer}}, \bibinfo {author}
  {\bibfnamefont {M.}~\bibnamefont {Dragicevic}}, \bibinfo {author}
  {\bibfnamefont {J.}~\bibnamefont {Er{\"o}}}, \bibinfo {author} {\bibfnamefont
  {C.}~\bibnamefont {Fabjan}}, \bibinfo {author} {\bibfnamefont
  {M.}~\bibnamefont {Friedl}},  \emph {et~al.},\ }\href {\doibase
  10.1007/JHEP06(2013)081} {\bibfield  {journal} {\bibinfo  {journal} {J. High
  Energ. Phys.}\ }\textbf {\bibinfo {volume} {2013}},\ \bibinfo {pages} {81}
  (\bibinfo {year} {2013})}\BibitemShut {NoStop}%
\bibitem [{\citenamefont {Zyla~\textit{et. al.} (Particle
  Data~Group)}(2020)}]{zyla2020review}%
  \BibitemOpen
  \bibfield  {author} {\bibinfo {author} {\bibfnamefont {P.~A.}\ \bibnamefont
  {Zyla~\textit{et. al.} (Particle Data~Group)}},\ }\href {\doibase
  10.1093/ptep/ptaa104} {\bibfield  {journal} {\bibinfo  {journal} {Prog.
  Theor. Exp. Phys.}\ }\textbf {\bibinfo {volume} {2020}},\ \bibinfo {pages}
  {083C01} (\bibinfo {year} {2020})}\BibitemShut {NoStop}%
\bibitem [{\citenamefont {Sakharov}(1967)}]{sakharov1967violation}%
  \BibitemOpen
  \bibfield  {author} {\bibinfo {author} {\bibfnamefont {A.~D.}\ \bibnamefont
  {Sakharov}},\ }\href {\doibase 10.1070/PU1991v034n05ABEH002497} {\bibfield
  {journal} {\bibinfo  {journal} {JETP Lett.}\ }\textbf {\bibinfo {volume}
  {5}},\ \bibinfo {pages} {24} (\bibinfo {year} {1967})}\BibitemShut {NoStop}%
\bibitem [{\citenamefont {Cerdeno}\ \emph {et~al.}(2020)\citenamefont
  {Cerdeno}, \citenamefont {Reimitz}, \citenamefont {Sakurai} \emph
  {et~al.}}]{cerdeno2020impact}%
  \BibitemOpen
  \bibfield  {author} {\bibinfo {author} {\bibfnamefont {D.~G.}\ \bibnamefont
  {Cerdeno}}, \bibinfo {author} {\bibfnamefont {P.}~\bibnamefont {Reimitz}},
  \bibinfo {author} {\bibfnamefont {K.}~\bibnamefont {Sakurai}},  \emph
  {et~al.},\ }\bibfield  {booktitle} {\emph {\bibinfo {booktitle} {J. Phys.:
  Conf. Ser.}},\ }\href {\doibase 10.1088/1742-6596/1586/1/012044} {\ \textbf
  {\bibinfo {volume} {1586}},\ \bibinfo {pages} {012044} (\bibinfo {year}
  {2020})}\BibitemShut {NoStop}%
\bibitem [{\citenamefont {'t~Hooft}(1976{\natexlab{a}})}]{hooft1976symmetry}%
  \BibitemOpen
  \bibfield  {author} {\bibinfo {author} {\bibfnamefont {G.}~\bibnamefont
  {'t~Hooft}},\ }\href {\doibase 10.1103/PhysRevLett.37.8} {\bibfield
  {journal} {\bibinfo  {journal} {Phys. Rev. Lett.}\ }\textbf {\bibinfo
  {volume} {37}},\ \bibinfo {pages} {8} (\bibinfo {year}
  {1976}{\natexlab{a}})}\BibitemShut {NoStop}%
\bibitem [{\citenamefont
  {'t~Hooft}(1976{\natexlab{b}})}]{hooft1976computation}%
  \BibitemOpen
  \bibfield  {author} {\bibinfo {author} {\bibfnamefont {G.}~\bibnamefont
  {'t~Hooft}},\ }\href {\doibase 10.1103/PhysRevD.14.3432} {\bibfield
  {journal} {\bibinfo  {journal} {Phys. Rev. D}\ }\textbf {\bibinfo {volume}
  {14}},\ \bibinfo {pages} {3432} (\bibinfo {year}
  {1976}{\natexlab{b}})}\BibitemShut {NoStop}%
\bibitem [{\citenamefont {Arnold}\ \emph {et~al.}(2013)\citenamefont {Arnold},
  \citenamefont {Fornal},\ and\ \citenamefont {Wise}}]{arnold2013simplified}%
  \BibitemOpen
  \bibfield  {author} {\bibinfo {author} {\bibfnamefont {J.~M.}\ \bibnamefont
  {Arnold}}, \bibinfo {author} {\bibfnamefont {B.}~\bibnamefont {Fornal}}, \
  and\ \bibinfo {author} {\bibfnamefont {M.~B.}\ \bibnamefont {Wise}},\ }\href
  {\doibase 10.1103/PhysRevD.87.075004} {\bibfield  {journal} {\bibinfo
  {journal} {Phys. Rev. D}\ }\textbf {\bibinfo {volume} {87}},\ \bibinfo
  {pages} {075004} (\bibinfo {year} {2013})}\BibitemShut {NoStop}%
\bibitem [{\citenamefont {Ellis}\ and\ \citenamefont
  {Sakurai}(2016)}]{ellis2016search}%
  \BibitemOpen
  \bibfield  {author} {\bibinfo {author} {\bibfnamefont {J.}~\bibnamefont
  {Ellis}}\ and\ \bibinfo {author} {\bibfnamefont {K.}~\bibnamefont
  {Sakurai}},\ }\href {\doibase 10.1007/JHEP04(2016)086} {\bibfield  {journal}
  {\bibinfo  {journal} {J. High Energ. Phys.}\ }\textbf {\bibinfo {volume}
  {2016}},\ \bibinfo {pages} {86} (\bibinfo {year} {2016})}\BibitemShut
  {NoStop}%
\bibitem [{\citenamefont {Kuzmin}\ \emph {et~al.}(1985)\citenamefont {Kuzmin},
  \citenamefont {Rubakov},\ and\ \citenamefont
  {Shaposhnikov}}]{kuzmin1985anomalous}%
  \BibitemOpen
  \bibfield  {author} {\bibinfo {author} {\bibfnamefont {V.~A.}\ \bibnamefont
  {Kuzmin}}, \bibinfo {author} {\bibfnamefont {V.~A.}\ \bibnamefont {Rubakov}},
  \ and\ \bibinfo {author} {\bibfnamefont {M.~E.}\ \bibnamefont
  {Shaposhnikov}},\ }\href {\doibase 10.1016/0370-2693(85)91028-7} {\bibfield
  {journal} {\bibinfo  {journal} {Phys. Lett. B}\ }\textbf {\bibinfo {volume}
  {155}},\ \bibinfo {pages} {36} (\bibinfo {year} {1985})}\BibitemShut
  {NoStop}%
\bibitem [{\citenamefont {Dev}\ \emph {et~al.}(2022)\citenamefont {Dev},
  \citenamefont {Koerner}, \citenamefont {Saad}, \citenamefont {Antusch},
  \citenamefont {Askins}, \citenamefont {Babu}, \citenamefont {Barrow},
  \citenamefont {Chakrabortty}, \citenamefont {de~Gouv{\^e}a}, \citenamefont
  {Djurcic} \emph {et~al.}}]{dev2022searches}%
  \BibitemOpen
  \bibfield  {author} {\bibinfo {author} {\bibfnamefont {P.~S.~B.}\
  \bibnamefont {Dev}}, \bibinfo {author} {\bibfnamefont {L.~W.}\ \bibnamefont
  {Koerner}}, \bibinfo {author} {\bibfnamefont {S.}~\bibnamefont {Saad}},
  \bibinfo {author} {\bibfnamefont {S.}~\bibnamefont {Antusch}}, \bibinfo
  {author} {\bibfnamefont {M.}~\bibnamefont {Askins}}, \bibinfo {author}
  {\bibfnamefont {K.~S.}\ \bibnamefont {Babu}}, \bibinfo {author}
  {\bibfnamefont {J.~L.}\ \bibnamefont {Barrow}}, \bibinfo {author}
  {\bibfnamefont {J.}~\bibnamefont {Chakrabortty}}, \bibinfo {author}
  {\bibfnamefont {A.}~\bibnamefont {de~Gouv{\^e}a}}, \bibinfo {author}
  {\bibfnamefont {Z.}~\bibnamefont {Djurcic}},  \emph {et~al.},\ }\href
  {https://arxiv.org/abs/2203.08771} {\bibfield  {journal} {\bibinfo  {journal}
  {arXiv:2203.08771}\ } (\bibinfo {year} {2022})}\BibitemShut {NoStop}%
\bibitem [{\citenamefont {Landau}(1957)}]{landau1957conservation}%
  \BibitemOpen
  \bibfield  {author} {\bibinfo {author} {\bibfnamefont {L.}~\bibnamefont
  {Landau}},\ }\href {\doibase 10.1016/0029-5582(57)90061-5} {\bibfield
  {journal} {\bibinfo  {journal} {Nucl. Phys.}\ }\textbf {\bibinfo {volume}
  {3}},\ \bibinfo {pages} {127} (\bibinfo {year} {1957})}\BibitemShut {NoStop}%
\bibitem [{\citenamefont {Okubo}(1958)}]{okubo1958decay}%
  \BibitemOpen
  \bibfield  {author} {\bibinfo {author} {\bibfnamefont {S.}~\bibnamefont
  {Okubo}},\ }\href {\doibase 10.1103/PhysRev.109.984} {\bibfield  {journal}
  {\bibinfo  {journal} {Phys. Rev.}\ }\textbf {\bibinfo {volume} {109}},\
  \bibinfo {pages} {984} (\bibinfo {year} {1958})}\BibitemShut {NoStop}%
\bibitem [{\citenamefont {Christenson}\ \emph {et~al.}(1964)\citenamefont
  {Christenson}, \citenamefont {Cronin}, \citenamefont {Fitch},\ and\
  \citenamefont {Turlay}}]{christenson1964evidence}%
  \BibitemOpen
  \bibfield  {author} {\bibinfo {author} {\bibfnamefont {J.~H.}\ \bibnamefont
  {Christenson}}, \bibinfo {author} {\bibfnamefont {J.~W.}\ \bibnamefont
  {Cronin}}, \bibinfo {author} {\bibfnamefont {V.~L.}\ \bibnamefont {Fitch}}, \
  and\ \bibinfo {author} {\bibfnamefont {R.}~\bibnamefont {Turlay}},\ }\href
  {\doibase 10.1103/PhysRevLett.13.138} {\bibfield  {journal} {\bibinfo
  {journal} {Phys. Rev. Lett.}\ }\textbf {\bibinfo {volume} {13}},\ \bibinfo
  {pages} {138} (\bibinfo {year} {1964})}\BibitemShut {NoStop}%
\bibitem [{\citenamefont {Aubert}\ \emph {et~al.}(2001)\citenamefont {Aubert},
  \citenamefont {Boutigny}, \citenamefont {Gaillard}, \citenamefont {Hicheur},
  \citenamefont {Karyotakis}, \citenamefont {Lees}, \citenamefont {Robbe},
  \citenamefont {Tisserand}, \citenamefont {Palano}, \citenamefont {Chen} \emph
  {et~al.}}]{aubert2001observation}%
  \BibitemOpen
  \bibfield  {author} {\bibinfo {author} {\bibfnamefont {B.}~\bibnamefont
  {Aubert}}, \bibinfo {author} {\bibfnamefont {D.}~\bibnamefont {Boutigny}},
  \bibinfo {author} {\bibfnamefont {J.~M.}\ \bibnamefont {Gaillard}}, \bibinfo
  {author} {\bibfnamefont {A.}~\bibnamefont {Hicheur}}, \bibinfo {author}
  {\bibfnamefont {Y.}~\bibnamefont {Karyotakis}}, \bibinfo {author}
  {\bibfnamefont {J.~P.}\ \bibnamefont {Lees}}, \bibinfo {author}
  {\bibfnamefont {P.}~\bibnamefont {Robbe}}, \bibinfo {author} {\bibfnamefont
  {V.}~\bibnamefont {Tisserand}}, \bibinfo {author} {\bibfnamefont
  {A.}~\bibnamefont {Palano}}, \bibinfo {author} {\bibfnamefont {G.~P.}\
  \bibnamefont {Chen}},  \emph {et~al.},\ }\href {\doibase
  10.1103/PhysRevLett.87.091801} {\bibfield  {journal} {\bibinfo  {journal}
  {Phys. Rev. Lett.}\ }\textbf {\bibinfo {volume} {87}},\ \bibinfo {pages}
  {091801} (\bibinfo {year} {2001})}\BibitemShut {NoStop}%
\bibitem [{\citenamefont {Abe}\ \emph {et~al.}(2001)\citenamefont {Abe},
  \citenamefont {Abe}, \citenamefont {Adachi}, \citenamefont {Ahn},
  \citenamefont {Aihara}, \citenamefont {Akatsu}, \citenamefont {Alimonti},
  \citenamefont {Asai}, \citenamefont {Asai}, \citenamefont {Asano} \emph
  {et~al.}}]{abe2001observation}%
  \BibitemOpen
  \bibfield  {author} {\bibinfo {author} {\bibfnamefont {K.}~\bibnamefont
  {Abe}}, \bibinfo {author} {\bibfnamefont {R.}~\bibnamefont {Abe}}, \bibinfo
  {author} {\bibfnamefont {I.}~\bibnamefont {Adachi}}, \bibinfo {author}
  {\bibfnamefont {B.~S.}\ \bibnamefont {Ahn}}, \bibinfo {author} {\bibfnamefont
  {H.}~\bibnamefont {Aihara}}, \bibinfo {author} {\bibfnamefont
  {M.}~\bibnamefont {Akatsu}}, \bibinfo {author} {\bibfnamefont
  {G.}~\bibnamefont {Alimonti}}, \bibinfo {author} {\bibfnamefont
  {K.}~\bibnamefont {Asai}}, \bibinfo {author} {\bibfnamefont {M.}~\bibnamefont
  {Asai}}, \bibinfo {author} {\bibfnamefont {Y.}~\bibnamefont {Asano}},  \emph
  {et~al.},\ }\href {\doibase 10.1103/PhysRevLett.87.091802} {\bibfield
  {journal} {\bibinfo  {journal} {Phys. Rev. Lett.}\ }\textbf {\bibinfo
  {volume} {87}},\ \bibinfo {pages} {091802} (\bibinfo {year}
  {2001})}\BibitemShut {NoStop}%
\bibitem [{\citenamefont {Aaij}\ \emph {et~al.}(2013)\citenamefont {Aaij},
  \citenamefont {Beteta}, \citenamefont {Adeva}, \citenamefont {Adinolfi},
  \citenamefont {Adrover}, \citenamefont {Affolder}, \citenamefont {Ajaltouni},
  \citenamefont {Albrecht}, \citenamefont {Alessio}, \citenamefont {Alexander}
  \emph {et~al.}}]{aaij2013first}%
  \BibitemOpen
  \bibfield  {author} {\bibinfo {author} {\bibfnamefont {R.}~\bibnamefont
  {Aaij}}, \bibinfo {author} {\bibfnamefont {C.~A.}\ \bibnamefont {Beteta}},
  \bibinfo {author} {\bibfnamefont {B.}~\bibnamefont {Adeva}}, \bibinfo
  {author} {\bibfnamefont {M.}~\bibnamefont {Adinolfi}}, \bibinfo {author}
  {\bibfnamefont {C.}~\bibnamefont {Adrover}}, \bibinfo {author} {\bibfnamefont
  {A.}~\bibnamefont {Affolder}}, \bibinfo {author} {\bibfnamefont
  {Z.}~\bibnamefont {Ajaltouni}}, \bibinfo {author} {\bibfnamefont
  {J.}~\bibnamefont {Albrecht}}, \bibinfo {author} {\bibfnamefont
  {F.}~\bibnamefont {Alessio}}, \bibinfo {author} {\bibfnamefont
  {M.}~\bibnamefont {Alexander}},  \emph {et~al.},\ }\href {\doibase
  10.1103/PhysRevLett.110.221601} {\bibfield  {journal} {\bibinfo  {journal}
  {Phys. Rev. Lett.}\ }\textbf {\bibinfo {volume} {110}},\ \bibinfo {pages}
  {221601} (\bibinfo {year} {2013})}\BibitemShut {NoStop}%
\bibitem [{\citenamefont {Aaij$}\ \emph {et~al.}(2019)\citenamefont {Aaij$}
  \emph {et~al.}}]{aaij2019observation}%
  \BibitemOpen
  \bibfield  {author} {\bibinfo {author} {\bibfnamefont {R.}~\bibnamefont
  {Aaij$}, \bibfnamefont {$}} \emph {et~al.},\ }\href {\doibase
  10.1103/PhysRevLett.122.211803} {\bibfield  {journal} {\bibinfo  {journal}
  {Phys. Rev. Lett.}\ }\textbf {\bibinfo {volume} {122}},\ \bibinfo {pages}
  {211803} (\bibinfo {year} {2019})}\BibitemShut {NoStop}%
\bibitem [{\citenamefont {Phillips~II}\ \emph {et~al.}(2016)\citenamefont
  {Phillips~II}, \citenamefont {Snow}, \citenamefont {Babu}, \citenamefont
  {Banerjee}, \citenamefont {Baxter}, \citenamefont {Berezhiani}, \citenamefont
  {Bergevin}, \citenamefont {Bhattacharya}, \citenamefont {Brooijmans},
  \citenamefont {Castellanos} \emph {et~al.}}]{phillips2016neutron}%
  \BibitemOpen
  \bibfield  {author} {\bibinfo {author} {\bibfnamefont {D.~G.}\ \bibnamefont
  {Phillips~II}}, \bibinfo {author} {\bibfnamefont {W.~M.}\ \bibnamefont
  {Snow}}, \bibinfo {author} {\bibfnamefont {K.}~\bibnamefont {Babu}}, \bibinfo
  {author} {\bibfnamefont {S.}~\bibnamefont {Banerjee}}, \bibinfo {author}
  {\bibfnamefont {D.~V.}\ \bibnamefont {Baxter}}, \bibinfo {author}
  {\bibfnamefont {Z.}~\bibnamefont {Berezhiani}}, \bibinfo {author}
  {\bibfnamefont {M.}~\bibnamefont {Bergevin}}, \bibinfo {author}
  {\bibfnamefont {S.}~\bibnamefont {Bhattacharya}}, \bibinfo {author}
  {\bibfnamefont {G.}~\bibnamefont {Brooijmans}}, \bibinfo {author}
  {\bibfnamefont {L.}~\bibnamefont {Castellanos}},  \emph {et~al.},\ }\href
  {\doibase 10.1016/j.physrep.2015.11.001} {\bibfield  {journal} {\bibinfo
  {journal} {Phys. Rep.}\ }\textbf {\bibinfo {volume} {612}},\ \bibinfo {pages}
  {1} (\bibinfo {year} {2016})}\BibitemShut {NoStop}%
\bibitem [{\citenamefont {Hao}\ and\ \citenamefont
  {Ni}(2022)}]{hao2022neutron}%
  \BibitemOpen
  \bibfield  {author} {\bibinfo {author} {\bibfnamefont {Y.}~\bibnamefont
  {Hao}}\ and\ \bibinfo {author} {\bibfnamefont {D.}~\bibnamefont {Ni}},\
  }\href {\doibase 10.1103/PhysRevD.106.115028} {\bibfield  {journal} {\bibinfo
   {journal} {Phys. Rev. D}\ }\textbf {\bibinfo {volume} {106}},\ \bibinfo
  {pages} {115028} (\bibinfo {year} {2022})}\BibitemShut {NoStop}%
\bibitem [{\citenamefont {Snow}\ \emph {et~al.}(2022)\citenamefont {Snow},
  \citenamefont {Haddock},\ and\ \citenamefont {Heacock}}]{snow2022searches}%
  \BibitemOpen
  \bibfield  {author} {\bibinfo {author} {\bibfnamefont {W.~M.}\ \bibnamefont
  {Snow}}, \bibinfo {author} {\bibfnamefont {C.}~\bibnamefont {Haddock}}, \
  and\ \bibinfo {author} {\bibfnamefont {B.}~\bibnamefont {Heacock}},\ }\href
  {\doibase 10.3390/sym14010010} {\bibfield  {journal} {\bibinfo  {journal}
  {Symm.}\ }\textbf {\bibinfo {volume} {14}},\ \bibinfo {pages} {10} (\bibinfo
  {year} {2022})}\BibitemShut {NoStop}%
\bibitem [{\citenamefont {Baldo-Ceolin}\ \emph {et~al.}(1994)\citenamefont
  {Baldo-Ceolin}, \citenamefont {Benetti}, \citenamefont {Bitter},
  \citenamefont {Bobisut}, \citenamefont {Calligarich}, \citenamefont
  {Dolfini}, \citenamefont {Dubbers}, \citenamefont {El-Muzeini}, \citenamefont
  {Genoni}, \citenamefont {Gibin} \emph {et~al.}}]{baldo1994new}%
  \BibitemOpen
  \bibfield  {author} {\bibinfo {author} {\bibfnamefont {M.}~\bibnamefont
  {Baldo-Ceolin}}, \bibinfo {author} {\bibfnamefont {P.}~\bibnamefont
  {Benetti}}, \bibinfo {author} {\bibfnamefont {T.}~\bibnamefont {Bitter}},
  \bibinfo {author} {\bibfnamefont {F.}~\bibnamefont {Bobisut}}, \bibinfo
  {author} {\bibfnamefont {E.}~\bibnamefont {Calligarich}}, \bibinfo {author}
  {\bibfnamefont {R.}~\bibnamefont {Dolfini}}, \bibinfo {author} {\bibfnamefont
  {D.}~\bibnamefont {Dubbers}}, \bibinfo {author} {\bibfnamefont
  {P.}~\bibnamefont {El-Muzeini}}, \bibinfo {author} {\bibfnamefont
  {M.}~\bibnamefont {Genoni}}, \bibinfo {author} {\bibfnamefont
  {D.}~\bibnamefont {Gibin}},  \emph {et~al.},\ }\href {\doibase
  10.1007/BF01580321} {\bibfield  {journal} {\bibinfo  {journal} {Z. Phys. C}\
  }\textbf {\bibinfo {volume} {63}},\ \bibinfo {pages} {409} (\bibinfo {year}
  {1994})}\BibitemShut {NoStop}%
\bibitem [{\citenamefont {Jones}\ \emph {et~al.}(1984)\citenamefont {Jones},
  \citenamefont {Bionta},\ and\ \citenamefont {Blewitt}}]{jones1984search}%
  \BibitemOpen
  \bibfield  {author} {\bibinfo {author} {\bibfnamefont {T.~W.}\ \bibnamefont
  {Jones}}, \bibinfo {author} {\bibfnamefont {R.~M.}\ \bibnamefont {Bionta}}, \
  and\ \bibinfo {author} {\bibfnamefont {G.}~\bibnamefont {Blewitt}},\ }\href
  {\doibase 10.1103/PhysRevLett.52.720} {\bibfield  {journal} {\bibinfo
  {journal} {Phys. Rev. Lett.}\ }\textbf {\bibinfo {volume} {52}},\ \bibinfo
  {pages} {720} (\bibinfo {year} {1984})}\BibitemShut {NoStop}%
\bibitem [{\citenamefont {Takita}\ \emph {et~al.}(1986)\citenamefont {Takita},
  \citenamefont {Arisaka}, \citenamefont {Kajita}, \citenamefont {Kifune},
  \citenamefont {Koshiba}, \citenamefont {Miyano}, \citenamefont {Nakahata},
  \citenamefont {Oyama}, \citenamefont {Sato}, \citenamefont {Suda} \emph
  {et~al.}}]{takita1986search}%
  \BibitemOpen
  \bibfield  {author} {\bibinfo {author} {\bibfnamefont {M.}~\bibnamefont
  {Takita}}, \bibinfo {author} {\bibfnamefont {K.}~\bibnamefont {Arisaka}},
  \bibinfo {author} {\bibfnamefont {T.}~\bibnamefont {Kajita}}, \bibinfo
  {author} {\bibfnamefont {T.}~\bibnamefont {Kifune}}, \bibinfo {author}
  {\bibfnamefont {M.}~\bibnamefont {Koshiba}}, \bibinfo {author} {\bibfnamefont
  {K.}~\bibnamefont {Miyano}}, \bibinfo {author} {\bibfnamefont
  {M.}~\bibnamefont {Nakahata}}, \bibinfo {author} {\bibfnamefont
  {Y.}~\bibnamefont {Oyama}}, \bibinfo {author} {\bibfnamefont
  {N.}~\bibnamefont {Sato}}, \bibinfo {author} {\bibfnamefont {T.}~\bibnamefont
  {Suda}},  \emph {et~al.},\ }\href {\doibase 10.1103/PhysRevD.34.902}
  {\bibfield  {journal} {\bibinfo  {journal} {Phys. Rev. D}\ }\textbf {\bibinfo
  {volume} {34}},\ \bibinfo {pages} {902} (\bibinfo {year} {1986})}\BibitemShut
  {NoStop}%
\bibitem [{\citenamefont {Berger}\ \emph {et~al.}(1990)\citenamefont {Berger},
  \citenamefont {Fr{\"o}hlich}, \citenamefont {M{\"o}nch}, \citenamefont
  {Nisius}, \citenamefont {Raupach}, \citenamefont {Schleper}, \citenamefont
  {Benadjal}, \citenamefont {Blum}, \citenamefont {Bourdarios}, \citenamefont
  {Dudelzak} \emph {et~al.}}]{berger1990search}%
  \BibitemOpen
  \bibfield  {author} {\bibinfo {author} {\bibfnamefont {C.}~\bibnamefont
  {Berger}}, \bibinfo {author} {\bibfnamefont {M.}~\bibnamefont
  {Fr{\"o}hlich}}, \bibinfo {author} {\bibfnamefont {H.}~\bibnamefont
  {M{\"o}nch}}, \bibinfo {author} {\bibfnamefont {R.}~\bibnamefont {Nisius}},
  \bibinfo {author} {\bibfnamefont {F.}~\bibnamefont {Raupach}}, \bibinfo
  {author} {\bibfnamefont {P.}~\bibnamefont {Schleper}}, \bibinfo {author}
  {\bibfnamefont {Y.}~\bibnamefont {Benadjal}}, \bibinfo {author}
  {\bibfnamefont {D.}~\bibnamefont {Blum}}, \bibinfo {author} {\bibfnamefont
  {C.}~\bibnamefont {Bourdarios}}, \bibinfo {author} {\bibfnamefont
  {B.}~\bibnamefont {Dudelzak}},  \emph {et~al.},\ }\href {\doibase
  10.1016/0370-2693(90)90441-8} {\bibfield  {journal} {\bibinfo  {journal}
  {Phys. Lett. B}\ }\textbf {\bibinfo {volume} {240}},\ \bibinfo {pages} {237}
  (\bibinfo {year} {1990})}\BibitemShut {NoStop}%
\bibitem [{\citenamefont {Chung}\ \emph {et~al.}(2002)\citenamefont {Chung},
  \citenamefont {Allison}, \citenamefont {Alner}, \citenamefont {Ayres},
  \citenamefont {Barrett}, \citenamefont {Border}, \citenamefont {Cobb},
  \citenamefont {Courant}, \citenamefont {Demuth}, \citenamefont {Fields} \emph
  {et~al.}}]{chung2002search}%
  \BibitemOpen
  \bibfield  {author} {\bibinfo {author} {\bibfnamefont {J.}~\bibnamefont
  {Chung}}, \bibinfo {author} {\bibfnamefont {W.}~\bibnamefont {Allison}},
  \bibinfo {author} {\bibfnamefont {G.~J.}\ \bibnamefont {Alner}}, \bibinfo
  {author} {\bibfnamefont {D.~S.}\ \bibnamefont {Ayres}}, \bibinfo {author}
  {\bibfnamefont {W.~L.}\ \bibnamefont {Barrett}}, \bibinfo {author}
  {\bibfnamefont {P.~M.}\ \bibnamefont {Border}}, \bibinfo {author}
  {\bibfnamefont {J.~H.}\ \bibnamefont {Cobb}}, \bibinfo {author}
  {\bibfnamefont {H.}~\bibnamefont {Courant}}, \bibinfo {author} {\bibfnamefont
  {D.~M.}\ \bibnamefont {Demuth}}, \bibinfo {author} {\bibfnamefont {T.~H.}\
  \bibnamefont {Fields}},  \emph {et~al.},\ }\href {\doibase
  10.1103/PhysRevD.66.032004} {\bibfield  {journal} {\bibinfo  {journal} {Phys.
  Rev. D}\ }\textbf {\bibinfo {volume} {66}},\ \bibinfo {pages} {032004}
  (\bibinfo {year} {2002})}\BibitemShut {NoStop}%
\bibitem [{\citenamefont {Aharmim}\ \emph {et~al.}(2017)\citenamefont
  {Aharmim}, \citenamefont {Ahmed}, \citenamefont {Anthony}, \citenamefont
  {Barros}, \citenamefont {Beier}, \citenamefont {Bellerive}, \citenamefont
  {Beltran}, \citenamefont {Bergevin}, \citenamefont {Biller}, \citenamefont
  {Boudjemline} \emph {et~al.}}]{aharmim2017search}%
  \BibitemOpen
  \bibfield  {author} {\bibinfo {author} {\bibfnamefont {B.}~\bibnamefont
  {Aharmim}}, \bibinfo {author} {\bibfnamefont {S.~N.}\ \bibnamefont {Ahmed}},
  \bibinfo {author} {\bibfnamefont {A.~E.}\ \bibnamefont {Anthony}}, \bibinfo
  {author} {\bibfnamefont {N.}~\bibnamefont {Barros}}, \bibinfo {author}
  {\bibfnamefont {E.~W.}\ \bibnamefont {Beier}}, \bibinfo {author}
  {\bibfnamefont {A.}~\bibnamefont {Bellerive}}, \bibinfo {author}
  {\bibfnamefont {B.}~\bibnamefont {Beltran}}, \bibinfo {author} {\bibfnamefont
  {M.}~\bibnamefont {Bergevin}}, \bibinfo {author} {\bibfnamefont {S.~D.}\
  \bibnamefont {Biller}}, \bibinfo {author} {\bibfnamefont {K.}~\bibnamefont
  {Boudjemline}},  \emph {et~al.},\ }\href {\doibase
  10.1103/PhysRevD.96.092005} {\bibfield  {journal} {\bibinfo  {journal} {Phys.
  Rev. D}\ }\textbf {\bibinfo {volume} {96}},\ \bibinfo {pages} {092005}
  (\bibinfo {year} {2017})}\BibitemShut {NoStop}%
\bibitem [{\citenamefont {Abe}\ \emph {et~al.}(2015)\citenamefont {Abe},
  \citenamefont {Hayato}, \citenamefont {Iida}, \citenamefont {Ishihara},
  \citenamefont {Kameda}, \citenamefont {Koshio}, \citenamefont {Minamino},
  \citenamefont {Mitsuda}, \citenamefont {Miura}, \citenamefont {Moriyama}
  \emph {et~al.}}]{abe2015search}%
  \BibitemOpen
  \bibfield  {author} {\bibinfo {author} {\bibfnamefont {K.}~\bibnamefont
  {Abe}}, \bibinfo {author} {\bibfnamefont {Y.}~\bibnamefont {Hayato}},
  \bibinfo {author} {\bibfnamefont {T.}~\bibnamefont {Iida}}, \bibinfo {author}
  {\bibfnamefont {K.}~\bibnamefont {Ishihara}}, \bibinfo {author}
  {\bibfnamefont {J.}~\bibnamefont {Kameda}}, \bibinfo {author} {\bibfnamefont
  {Y.}~\bibnamefont {Koshio}}, \bibinfo {author} {\bibfnamefont
  {A.}~\bibnamefont {Minamino}}, \bibinfo {author} {\bibfnamefont
  {C.}~\bibnamefont {Mitsuda}}, \bibinfo {author} {\bibfnamefont
  {M.}~\bibnamefont {Miura}}, \bibinfo {author} {\bibfnamefont
  {S.}~\bibnamefont {Moriyama}},  \emph {et~al.},\ }\href {\doibase
  10.1103/PhysRevD.91.072006} {\bibfield  {journal} {\bibinfo  {journal} {Phys.
  Rev. D}\ }\textbf {\bibinfo {volume} {91}},\ \bibinfo {pages} {072006}
  (\bibinfo {year} {2015})}\BibitemShut {NoStop}%
\bibitem [{\citenamefont {Abe}\ \emph {et~al.}(2021)\citenamefont {Abe},
  \citenamefont {Bronner}, \citenamefont {Hayato}, \citenamefont {Ikeda},
  \citenamefont {Imaizumi}, \citenamefont {Ito}, \citenamefont {Kameda},
  \citenamefont {Kataoka}, \citenamefont {Miura}, \citenamefont {Moriyama}
  \emph {et~al.}}]{abe2021neutron}%
  \BibitemOpen
  \bibfield  {author} {\bibinfo {author} {\bibfnamefont {K.}~\bibnamefont
  {Abe}}, \bibinfo {author} {\bibfnamefont {C.}~\bibnamefont {Bronner}},
  \bibinfo {author} {\bibfnamefont {Y.}~\bibnamefont {Hayato}}, \bibinfo
  {author} {\bibfnamefont {M.}~\bibnamefont {Ikeda}}, \bibinfo {author}
  {\bibfnamefont {S.}~\bibnamefont {Imaizumi}}, \bibinfo {author}
  {\bibfnamefont {H.}~\bibnamefont {Ito}}, \bibinfo {author} {\bibfnamefont
  {J.}~\bibnamefont {Kameda}}, \bibinfo {author} {\bibfnamefont
  {Y.}~\bibnamefont {Kataoka}}, \bibinfo {author} {\bibfnamefont
  {M.}~\bibnamefont {Miura}}, \bibinfo {author} {\bibfnamefont
  {S.}~\bibnamefont {Moriyama}},  \emph {et~al.},\ }\href {\doibase
  10.1103/PhysRevD.103.012008} {\bibfield  {journal} {\bibinfo  {journal}
  {Phys. Rev. D}\ }\textbf {\bibinfo {volume} {103}},\ \bibinfo {pages}
  {012008} (\bibinfo {year} {2021})}\BibitemShut {NoStop}%
\bibitem [{\citenamefont {Mohapatra}\ and\ \citenamefont
  {Okada}(2021)}]{mohapatra2021affleck}%
  \BibitemOpen
  \bibfield  {author} {\bibinfo {author} {\bibfnamefont {R.~N.}\ \bibnamefont
  {Mohapatra}}\ and\ \bibinfo {author} {\bibfnamefont {N.}~\bibnamefont
  {Okada}},\ }\href {\doibase 10.1103/PhysRevD.104.055030} {\bibfield
  {journal} {\bibinfo  {journal} {Phys. Rev. D}\ }\textbf {\bibinfo {volume}
  {104}},\ \bibinfo {pages} {055030} (\bibinfo {year} {2021})}\BibitemShut
  {NoStop}%
\bibitem [{\citenamefont {Affleck}\ and\ \citenamefont
  {Dine}(1985)}]{affleck1985new}%
  \BibitemOpen
  \bibfield  {author} {\bibinfo {author} {\bibfnamefont {I.}~\bibnamefont
  {Affleck}}\ and\ \bibinfo {author} {\bibfnamefont {M.}~\bibnamefont {Dine}},\
  }\href {\doibase 10.1016/0550-3213(85)90021-5} {\bibfield  {journal}
  {\bibinfo  {journal} {Nucl. Phys. B}\ }\textbf {\bibinfo {volume} {249}},\
  \bibinfo {pages} {361} (\bibinfo {year} {1985})}\BibitemShut {NoStop}%
\bibitem [{\citenamefont {Dine}\ \emph {et~al.}(1995)\citenamefont {Dine},
  \citenamefont {Randall},\ and\ \citenamefont
  {Thomas}}]{dine1995supersymmetry}%
  \BibitemOpen
  \bibfield  {author} {\bibinfo {author} {\bibfnamefont {M.}~\bibnamefont
  {Dine}}, \bibinfo {author} {\bibfnamefont {L.}~\bibnamefont {Randall}}, \
  and\ \bibinfo {author} {\bibfnamefont {S.}~\bibnamefont {Thomas}},\ }\href
  {\doibase 10.1103/PhysRevLett.75.398} {\bibfield  {journal} {\bibinfo
  {journal} {Phys. Rev. Lett.}\ }\textbf {\bibinfo {volume} {75}},\ \bibinfo
  {pages} {398} (\bibinfo {year} {1995})}\BibitemShut {NoStop}%
\bibitem [{\citenamefont {Dine}\ \emph {et~al.}(1996)\citenamefont {Dine},
  \citenamefont {Randall},\ and\ \citenamefont
  {Thomas}}]{dine1996baryogenesis}%
  \BibitemOpen
  \bibfield  {author} {\bibinfo {author} {\bibfnamefont {M.}~\bibnamefont
  {Dine}}, \bibinfo {author} {\bibfnamefont {L.}~\bibnamefont {Randall}}, \
  and\ \bibinfo {author} {\bibfnamefont {S.}~\bibnamefont {Thomas}},\ }\href
  {\doibase 10.1016/0550-3213(95)00538-2} {\bibfield  {journal} {\bibinfo
  {journal} {Nucl. Phys. B}\ }\textbf {\bibinfo {volume} {458}},\ \bibinfo
  {pages} {291} (\bibinfo {year} {1996})}\BibitemShut {NoStop}%
\bibitem [{\citenamefont {Babichev}\ \emph {et~al.}(2019)\citenamefont
  {Babichev}, \citenamefont {Gorbunov},\ and\ \citenamefont
  {Ramazanov}}]{babichev2019affleck}%
  \BibitemOpen
  \bibfield  {author} {\bibinfo {author} {\bibfnamefont {E.}~\bibnamefont
  {Babichev}}, \bibinfo {author} {\bibfnamefont {D.}~\bibnamefont {Gorbunov}},
  \ and\ \bibinfo {author} {\bibfnamefont {S.}~\bibnamefont {Ramazanov}},\
  }\href {\doibase 10.1016/j.physletb.2019.03.046} {\bibfield  {journal}
  {\bibinfo  {journal} {Phys. Lett. B}\ }\textbf {\bibinfo {volume} {792}},\
  \bibinfo {pages} {228} (\bibinfo {year} {2019})}\BibitemShut {NoStop}%
\bibitem [{\citenamefont {Lloyd-Stubbs}\ and\ \citenamefont
  {McDonald}(2021)}]{lloyd2021minimal}%
  \BibitemOpen
  \bibfield  {author} {\bibinfo {author} {\bibfnamefont {A.}~\bibnamefont
  {Lloyd-Stubbs}}\ and\ \bibinfo {author} {\bibfnamefont {J.}~\bibnamefont
  {McDonald}},\ }\href {\doibase 10.1103/PhysRevD.103.123514} {\bibfield
  {journal} {\bibinfo  {journal} {Phys. Rev. D}\ }\textbf {\bibinfo {volume}
  {103}},\ \bibinfo {pages} {123514} (\bibinfo {year} {2021})}\BibitemShut
  {NoStop}%
\bibitem [{\citenamefont {Migdal}(1959)}]{migdal1959superfluidity}%
  \BibitemOpen
  \bibfield  {author} {\bibinfo {author} {\bibfnamefont {A.~B.}\ \bibnamefont
  {Migdal}},\ }\href {\doibase 10.1016/0029-5582(59)90264-0} {\bibfield
  {journal} {\bibinfo  {journal} {Nucl. Phys.}\ }\textbf {\bibinfo {volume}
  {13}},\ \bibinfo {pages} {655} (\bibinfo {year} {1959})}\BibitemShut
  {NoStop}%
\bibitem [{\citenamefont {Ginzburg}\ and\ \citenamefont
  {Landau}(1950)}]{ginzburg1950on}%
  \BibitemOpen
  \bibfield  {author} {\bibinfo {author} {\bibfnamefont {V.~L.}\ \bibnamefont
  {Ginzburg}}\ and\ \bibinfo {author} {\bibfnamefont {L.~D.}\ \bibnamefont
  {Landau}},\ }\href {\doibase 10.1016/B978-0-08-010586-4.50035-3} {\bibfield
  {journal} {\bibinfo  {journal} {Zh. Eksp. Teor. Fiz.}\ }\textbf {\bibinfo
  {volume} {20}},\ \bibinfo {pages} {1064} (\bibinfo {year}
  {1950})}\BibitemShut {NoStop}%
\bibitem [{\citenamefont {Ginzburg}(2009)}]{ginzburg2009superconductivity}%
  \BibitemOpen
  \bibfield  {author} {\bibinfo {author} {\bibfnamefont {V.~L.}\ \bibnamefont
  {Ginzburg}},\ }\href
  {https://link.springer.com/book/10.1007/978-3-540-68008-6} {\emph {\bibinfo
  {title} {On superconductivity and superfluidity: a scientific
  autobiography}}}\ (\bibinfo  {publisher} {Springer},\ \bibinfo {address}
  {Heidelberg},\ \bibinfo {year} {2009})\BibitemShut {NoStop}%
\bibitem [{\citenamefont {Englert}\ and\ \citenamefont
  {Brout}(1964)}]{englert1964broken}%
  \BibitemOpen
  \bibfield  {author} {\bibinfo {author} {\bibfnamefont {F.}~\bibnamefont
  {Englert}}\ and\ \bibinfo {author} {\bibfnamefont {R.}~\bibnamefont
  {Brout}},\ }\href {\doibase 10.1103/PhysRevLett.13.321} {\bibfield  {journal}
  {\bibinfo  {journal} {Phys. Rev. Lett.}\ }\textbf {\bibinfo {volume} {13}},\
  \bibinfo {pages} {321} (\bibinfo {year} {1964})}\BibitemShut {NoStop}%
\bibitem [{\citenamefont {Higgs}(1964{\natexlab{a}})}]{higgs1964broken1}%
  \BibitemOpen
  \bibfield  {author} {\bibinfo {author} {\bibfnamefont {P.~W.}\ \bibnamefont
  {Higgs}},\ }\href {\doibase 10.1103/PhysRevLett.13.508} {\bibfield  {journal}
  {\bibinfo  {journal} {Phys. Rev. Lett.}\ }\textbf {\bibinfo {volume} {13}},\
  \bibinfo {pages} {508} (\bibinfo {year} {1964}{\natexlab{a}})}\BibitemShut
  {NoStop}%
\bibitem [{\citenamefont {Higgs}(1964{\natexlab{b}})}]{higgs1964broken2}%
  \BibitemOpen
  \bibfield  {author} {\bibinfo {author} {\bibfnamefont {P.~W.}\ \bibnamefont
  {Higgs}},\ }\href {\doibase 10.1016/0031-9163(64)91136-9} {\bibfield
  {journal} {\bibinfo  {journal} {Phys. Lett.}\ }\textbf {\bibinfo {volume}
  {12}},\ \bibinfo {pages} {132} (\bibinfo {year}
  {1964}{\natexlab{b}})}\BibitemShut {NoStop}%
\bibitem [{\citenamefont {Guralnik}\ \emph {et~al.}(1964)\citenamefont
  {Guralnik}, \citenamefont {Hagen},\ and\ \citenamefont
  {Kibble}}]{guralnik1964global}%
  \BibitemOpen
  \bibfield  {author} {\bibinfo {author} {\bibfnamefont {G.~S.}\ \bibnamefont
  {Guralnik}}, \bibinfo {author} {\bibfnamefont {C.~R.}\ \bibnamefont {Hagen}},
  \ and\ \bibinfo {author} {\bibfnamefont {T.~W.~B.}\ \bibnamefont {Kibble}},\
  }\href {\doibase 10.1103/PhysRevLett.13.585} {\bibfield  {journal} {\bibinfo
  {journal} {Phys. Rev. Lett.}\ }\textbf {\bibinfo {volume} {13}},\ \bibinfo
  {pages} {585} (\bibinfo {year} {1964})}\BibitemShut {NoStop}%
\bibitem [{\citenamefont {Bardeen}\ \emph {et~al.}(1957)\citenamefont
  {Bardeen}, \citenamefont {Cooper},\ and\ \citenamefont
  {Schrieffer}}]{bardeen1957theory}%
  \BibitemOpen
  \bibfield  {author} {\bibinfo {author} {\bibfnamefont {J.}~\bibnamefont
  {Bardeen}}, \bibinfo {author} {\bibfnamefont {L.~N.}\ \bibnamefont {Cooper}},
  \ and\ \bibinfo {author} {\bibfnamefont {J.~R.}\ \bibnamefont {Schrieffer}},\
  }\href {\doibase 10.1103/PhysRev.108.1175} {\bibfield  {journal} {\bibinfo
  {journal} {Phys. Rev.}\ }\textbf {\bibinfo {volume} {108}},\ \bibinfo {pages}
  {1175} (\bibinfo {year} {1957})}\BibitemShut {NoStop}%
\bibitem [{\citenamefont {Berti}\ \emph {et~al.}(2023)\citenamefont {Berti},
  \citenamefont {Cominotti}, \citenamefont {Rogora}, \citenamefont {Moss},
  \citenamefont {Billam}, \citenamefont {Carusotto}, \citenamefont {Lamporesi},
  \citenamefont {Recati}, \citenamefont {Ferrari},\ and\ \citenamefont
  {Zenesini}}]{berti2023observation}%
  \BibitemOpen
  \bibfield  {author} {\bibinfo {author} {\bibfnamefont {A.}~\bibnamefont
  {Berti}}, \bibinfo {author} {\bibfnamefont {R.}~\bibnamefont {Cominotti}},
  \bibinfo {author} {\bibfnamefont {C.}~\bibnamefont {Rogora}}, \bibinfo
  {author} {\bibfnamefont {I.}~\bibnamefont {Moss}}, \bibinfo {author}
  {\bibfnamefont {T.}~\bibnamefont {Billam}}, \bibinfo {author} {\bibfnamefont
  {I.}~\bibnamefont {Carusotto}}, \bibinfo {author} {\bibfnamefont
  {G.}~\bibnamefont {Lamporesi}}, \bibinfo {author} {\bibfnamefont
  {A.}~\bibnamefont {Recati}}, \bibinfo {author} {\bibfnamefont
  {G.}~\bibnamefont {Ferrari}}, \ and\ \bibinfo {author} {\bibfnamefont
  {A.}~\bibnamefont {Zenesini}},\ }\href {https://arxiv.org/abs/2305.05225}
  {\bibfield  {journal} {\bibinfo  {journal} {arXiv:2305.05225}\ } (\bibinfo
  {year} {2023})}\BibitemShut {NoStop}%
\bibitem [{\citenamefont {Lagnese}\ \emph {et~al.}(2023)\citenamefont
  {Lagnese}, \citenamefont {Surace}, \citenamefont {Morampudi},\ and\
  \citenamefont {Wilczek}}]{lagnese2023detecting}%
  \BibitemOpen
  \bibfield  {author} {\bibinfo {author} {\bibfnamefont {G.}~\bibnamefont
  {Lagnese}}, \bibinfo {author} {\bibfnamefont {F.~M.}\ \bibnamefont {Surace}},
  \bibinfo {author} {\bibfnamefont {S.}~\bibnamefont {Morampudi}}, \ and\
  \bibinfo {author} {\bibfnamefont {F.}~\bibnamefont {Wilczek}},\ }\href
  {https://arxiv.org/abs/2308.08340} {\bibfield  {journal} {\bibinfo  {journal}
  {arXiv:2308.08340}\ } (\bibinfo {year} {2023})}\BibitemShut {NoStop}%
\bibitem [{\citenamefont {Brink}\ and\ \citenamefont
  {Broglia}(2005)}]{brink2005nuclear}%
  \BibitemOpen
  \bibfield  {author} {\bibinfo {author} {\bibfnamefont {D.~M.}\ \bibnamefont
  {Brink}}\ and\ \bibinfo {author} {\bibfnamefont {R.~A.}\ \bibnamefont
  {Broglia}},\ }\href {\doibase 10.1017/9781009401920} {\emph {\bibinfo {title}
  {Nuclear Superfluidity: pairing in finite systems}}}\ (\bibinfo  {publisher}
  {Cambridge University Press},\ \bibinfo {address} {Cambridge},\ \bibinfo
  {year} {2005})\BibitemShut {NoStop}%
\bibitem [{\citenamefont {Haensel}\ \emph {et~al.}(2007)\citenamefont
  {Haensel}, \citenamefont {Potekhin},\ and\ \citenamefont
  {Yakovlev}}]{haensel2007neutron}%
  \BibitemOpen
  \bibfield  {author} {\bibinfo {author} {\bibfnamefont {P.}~\bibnamefont
  {Haensel}}, \bibinfo {author} {\bibfnamefont {A.~Y.}\ \bibnamefont
  {Potekhin}}, \ and\ \bibinfo {author} {\bibfnamefont {D.~G.}\ \bibnamefont
  {Yakovlev}},\ }\href {\doibase 10.1007/978-0-387-47301-7} {\emph {\bibinfo
  {title} {Neutron Stars 1: Equation of State and Structure}}}\ (\bibinfo
  {publisher} {Springer},\ \bibinfo {address} {New York},\ \bibinfo {year}
  {2007})\BibitemShut {NoStop}%
\bibitem [{\citenamefont {Bohr}\ \emph {et~al.}(1958)\citenamefont {Bohr},
  \citenamefont {Mottelson},\ and\ \citenamefont {Pines}}]{bohr1958possible}%
  \BibitemOpen
  \bibfield  {author} {\bibinfo {author} {\bibfnamefont {A.}~\bibnamefont
  {Bohr}}, \bibinfo {author} {\bibfnamefont {B.~R.}\ \bibnamefont {Mottelson}},
  \ and\ \bibinfo {author} {\bibfnamefont {D.}~\bibnamefont {Pines}},\ }\href
  {\doibase 10.1103/PhysRev.110.936} {\bibfield  {journal} {\bibinfo  {journal}
  {Phys. Rev.}\ }\textbf {\bibinfo {volume} {110}},\ \bibinfo {pages} {936}
  (\bibinfo {year} {1958})}\BibitemShut {NoStop}%
\bibitem [{\citenamefont {Anderson}\ and\ \citenamefont
  {Itoh}(1975)}]{anderson1975pulsar}%
  \BibitemOpen
  \bibfield  {author} {\bibinfo {author} {\bibfnamefont {P.~W.}\ \bibnamefont
  {Anderson}}\ and\ \bibinfo {author} {\bibfnamefont {N.}~\bibnamefont
  {Itoh}},\ }\href {\doibase 10.1038/256025a0} {\bibfield  {journal} {\bibinfo
  {journal} {Nature}\ }\textbf {\bibinfo {volume} {256}},\ \bibinfo {pages}
  {25} (\bibinfo {year} {1975})}\BibitemShut {NoStop}%
\bibitem [{\citenamefont {Yakovlev}\ and\ \citenamefont
  {Pethick}(2004)}]{yakovlev2004neutron}%
  \BibitemOpen
  \bibfield  {author} {\bibinfo {author} {\bibfnamefont {D.~G.}\ \bibnamefont
  {Yakovlev}}\ and\ \bibinfo {author} {\bibfnamefont {C.~J.}\ \bibnamefont
  {Pethick}},\ }\href {\doibase 10.1146/annurev.astro.42.053102.134013}
  {\bibfield  {journal} {\bibinfo  {journal} {Annu. Rev. Astron. Astrophys.}\
  }\textbf {\bibinfo {volume} {42}},\ \bibinfo {pages} {169} (\bibinfo {year}
  {2004})}\BibitemShut {NoStop}%
\bibitem [{\citenamefont {Page}\ \emph {et~al.}(2004)\citenamefont {Page},
  \citenamefont {Lattimer}, \citenamefont {Prakash},\ and\ \citenamefont
  {Steiner}}]{page2004minimal}%
  \BibitemOpen
  \bibfield  {author} {\bibinfo {author} {\bibfnamefont {D.}~\bibnamefont
  {Page}}, \bibinfo {author} {\bibfnamefont {J.~M.}\ \bibnamefont {Lattimer}},
  \bibinfo {author} {\bibfnamefont {M.}~\bibnamefont {Prakash}}, \ and\
  \bibinfo {author} {\bibfnamefont {A.~W.}\ \bibnamefont {Steiner}},\ }\href
  {\doibase 10.1086/424844} {\bibfield  {journal} {\bibinfo  {journal}
  {Astrophys. J. Suppl. Ser.}\ }\textbf {\bibinfo {volume} {155}},\ \bibinfo
  {pages} {623} (\bibinfo {year} {2004})}\BibitemShut {NoStop}%
\bibitem [{\citenamefont {Page}\ \emph {et~al.}(2006)\citenamefont {Page},
  \citenamefont {Geppert},\ and\ \citenamefont {Weber}}]{page2006cooling}%
  \BibitemOpen
  \bibfield  {author} {\bibinfo {author} {\bibfnamefont {D.}~\bibnamefont
  {Page}}, \bibinfo {author} {\bibfnamefont {U.}~\bibnamefont {Geppert}}, \
  and\ \bibinfo {author} {\bibfnamefont {F.}~\bibnamefont {Weber}},\ }\href
  {\doibase 10.1016/j.nuclphysa.2005.09.019} {\bibfield  {journal} {\bibinfo
  {journal} {Nucl. Phys. A}\ }\textbf {\bibinfo {volume} {777}},\ \bibinfo
  {pages} {497} (\bibinfo {year} {2006})}\BibitemShut {NoStop}%
\bibitem [{\citenamefont {Oertel}\ \emph {et~al.}(2017)\citenamefont {Oertel},
  \citenamefont {Hempel}, \citenamefont {Kl{\"a}hn},\ and\ \citenamefont
  {Typel}}]{oertel2017equations}%
  \BibitemOpen
  \bibfield  {author} {\bibinfo {author} {\bibfnamefont {M.}~\bibnamefont
  {Oertel}}, \bibinfo {author} {\bibfnamefont {M.}~\bibnamefont {Hempel}},
  \bibinfo {author} {\bibfnamefont {T.}~\bibnamefont {Kl{\"a}hn}}, \ and\
  \bibinfo {author} {\bibfnamefont {S.}~\bibnamefont {Typel}},\ }\href
  {\doibase 10.1103/RevModPhys.89.015007} {\bibfield  {journal} {\bibinfo
  {journal} {Rev. Mod. Phys.}\ }\textbf {\bibinfo {volume} {89}},\ \bibinfo
  {pages} {015007} (\bibinfo {year} {2017})}\BibitemShut {NoStop}%
\bibitem [{\citenamefont {Potekhin}\ \emph {et~al.}(2015)\citenamefont
  {Potekhin}, \citenamefont {Pons},\ and\ \citenamefont
  {Page}}]{potekhin2015neutron}%
  \BibitemOpen
  \bibfield  {author} {\bibinfo {author} {\bibfnamefont {A.~Y.}\ \bibnamefont
  {Potekhin}}, \bibinfo {author} {\bibfnamefont {J.~A.}\ \bibnamefont {Pons}},
  \ and\ \bibinfo {author} {\bibfnamefont {D.}~\bibnamefont {Page}},\ }\href
  {\doibase 10.1007/s11214-015-0180-9} {\bibfield  {journal} {\bibinfo
  {journal} {Space Sci. Rev.}\ }\textbf {\bibinfo {volume} {191}},\ \bibinfo
  {pages} {239} (\bibinfo {year} {2015})}\BibitemShut {NoStop}%
\bibitem [{\citenamefont {Wolf}(1966)}]{wolf1966some}%
  \BibitemOpen
  \bibfield  {author} {\bibinfo {author} {\bibfnamefont {R.~A.}\ \bibnamefont
  {Wolf}},\ }\href {\doibase 10.1086/148829} {\bibfield  {journal} {\bibinfo
  {journal} {Astrophys. J.}\ }\textbf {\bibinfo {volume} {145}},\ \bibinfo
  {pages} {834} (\bibinfo {year} {1966})}\BibitemShut {NoStop}%
\bibitem [{\citenamefont {Ruderman}(1967)}]{ruderman1967states}%
  \BibitemOpen
  \bibfield  {author} {\bibinfo {author} {\bibfnamefont {M.~A.}\ \bibnamefont
  {Ruderman}},\ }in\ \href {https://www.osti.gov/biblio/4493633} {\emph
  {\bibinfo {booktitle} {Proceedings of the Fifth Annual Eastern Theoretical
  Physics Conference}}},\ \bibinfo {editor} {edited by\ \bibinfo {editor}
  {\bibfnamefont {D.}~\bibnamefont {Feldman}}}\ (\bibinfo  {publisher} {W. A.
  Benjamin},\ \bibinfo {address} {New York},\ \bibinfo {year}
  {1967})\BibitemShut {NoStop}%
\bibitem [{\citenamefont {Hoffberg}\ \emph {et~al.}(1970)\citenamefont
  {Hoffberg}, \citenamefont {Glassgold}, \citenamefont {Richardson},\ and\
  \citenamefont {Ruderman}}]{hoffberg1970anisotropic}%
  \BibitemOpen
  \bibfield  {author} {\bibinfo {author} {\bibfnamefont {M.}~\bibnamefont
  {Hoffberg}}, \bibinfo {author} {\bibfnamefont {A.~E.}\ \bibnamefont
  {Glassgold}}, \bibinfo {author} {\bibfnamefont {R.~W.}\ \bibnamefont
  {Richardson}}, \ and\ \bibinfo {author} {\bibfnamefont {M.}~\bibnamefont
  {Ruderman}},\ }\href {\doibase 10.1103/PhysRevLett.24.775} {\bibfield
  {journal} {\bibinfo  {journal} {Phys. Rev. Lett.}\ }\textbf {\bibinfo
  {volume} {24}},\ \bibinfo {pages} {775} (\bibinfo {year} {1970})}\BibitemShut
  {NoStop}%
\bibitem [{\citenamefont {Mereghetti}\ \emph {et~al.}(2015)\citenamefont
  {Mereghetti}, \citenamefont {Pons},\ and\ \citenamefont
  {Melatos}}]{mereghetti2015magnetars}%
  \BibitemOpen
  \bibfield  {author} {\bibinfo {author} {\bibfnamefont {S.}~\bibnamefont
  {Mereghetti}}, \bibinfo {author} {\bibfnamefont {J.~A.}\ \bibnamefont
  {Pons}}, \ and\ \bibinfo {author} {\bibfnamefont {A.}~\bibnamefont
  {Melatos}},\ }\href {\doibase 10.1007/s11214-015-0146-y} {\bibfield
  {journal} {\bibinfo  {journal} {Space Sci. Rev.}\ }\textbf {\bibinfo {volume}
  {191}},\ \bibinfo {pages} {315} (\bibinfo {year} {2015})}\BibitemShut
  {NoStop}%
\bibitem [{\citenamefont {Mangin}\ and\ \citenamefont
  {Kahn}(2017)}]{mangin2017superconductivity}%
  \BibitemOpen
  \bibfield  {author} {\bibinfo {author} {\bibfnamefont {P.}~\bibnamefont
  {Mangin}}\ and\ \bibinfo {author} {\bibfnamefont {R.}~\bibnamefont {Kahn}},\
  }\href {https://link.springer.com/book/10.1007/978-3-319-50527-5} {\emph
  {\bibinfo {title} {Superconductivity: An introduction}}}\ (\bibinfo
  {publisher} {Springer},\ \bibinfo {address} {Grenoble},\ \bibinfo {year}
  {2017})\BibitemShut {NoStop}%
\bibitem [{\citenamefont {Bjorken}\ \emph {et~al.}(1964)\citenamefont
  {Bjorken}, \citenamefont {Drell},\ and\ \citenamefont
  {Mansfield}}]{bjorken1964relativistic}%
  \BibitemOpen
  \bibfield  {author} {\bibinfo {author} {\bibfnamefont {J.~D.}\ \bibnamefont
  {Bjorken}}, \bibinfo {author} {\bibfnamefont {S.~D.}\ \bibnamefont {Drell}},
  \ and\ \bibinfo {author} {\bibfnamefont {J.~E.}\ \bibnamefont {Mansfield}},\
  }\href@noop {} {\emph {\bibinfo {title} {Relativistic quantum mechanics}}}\
  (\bibinfo  {publisher} {McGraw-Hill},\ \bibinfo {address} {New York},\
  \bibinfo {year} {1964})\BibitemShut {NoStop}%
\bibitem [{\citenamefont {Bentez}\ \emph
  {et~al.}(1990{\natexlab{a}})\citenamefont {Bentez}, \citenamefont {Martnez-y
  Romero}, \citenamefont {N{\'u}ez-Y{\'e}pez},\ and\ \citenamefont
  {Salas-Brito}}]{bentez1990solution}%
  \BibitemOpen
  \bibfield  {author} {\bibinfo {author} {\bibfnamefont {J.}~\bibnamefont
  {Bentez}}, \bibinfo {author} {\bibfnamefont {R.~P.}\ \bibnamefont {Martnez-y
  Romero}}, \bibinfo {author} {\bibfnamefont {H.~N.}\ \bibnamefont
  {N{\'u}ez-Y{\'e}pez}}, \ and\ \bibinfo {author} {\bibfnamefont {A.~L.}\
  \bibnamefont {Salas-Brito}},\ }\href {\doibase 10.1103/PhysRevLett.64.1643}
  {\bibfield  {journal} {\bibinfo  {journal} {Phys. Rev. Lett.}\ }\textbf
  {\bibinfo {volume} {64}},\ \bibinfo {pages} {1643} (\bibinfo {year}
  {1990}{\natexlab{a}})}\BibitemShut {NoStop}%
\bibitem [{\citenamefont {Bentez}\ \emph
  {et~al.}(1990{\natexlab{b}})\citenamefont {Bentez}, \citenamefont {Martnez-y
  Romero}, \citenamefont {N{\'u}ez-Y{\'e}pez},\ and\ \citenamefont
  {Salas-Brito}}]{benitez1990erratum}%
  \BibitemOpen
  \bibfield  {author} {\bibinfo {author} {\bibfnamefont {J.}~\bibnamefont
  {Bentez}}, \bibinfo {author} {\bibfnamefont {R.~P.}\ \bibnamefont {Martnez-y
  Romero}}, \bibinfo {author} {\bibfnamefont {H.~N.}\ \bibnamefont
  {N{\'u}ez-Y{\'e}pez}}, \ and\ \bibinfo {author} {\bibfnamefont {A.~L.}\
  \bibnamefont {Salas-Brito}},\ }\href {\doibase 10.1103/PhysRevLett.65.2085}
  {\bibfield  {journal} {\bibinfo  {journal} {Phys. Rev. Lett.}\ }\textbf
  {\bibinfo {volume} {65}},\ \bibinfo {pages} {2085} (\bibinfo {year}
  {1990}{\natexlab{b}})}\BibitemShut {NoStop}%
\bibitem [{\citenamefont {Saxena}(2012)}]{saxena2012high}%
  \BibitemOpen
  \bibfield  {author} {\bibinfo {author} {\bibfnamefont {A.~K.}\ \bibnamefont
  {Saxena}},\ }\href {https://link.springer.com/book/10.1007/978-3-642-28481-6}
  {\emph {\bibinfo {title} {High-temperature superconductors}}}\ (\bibinfo
  {publisher} {Springer},\ \bibinfo {address} {Berlin, Heidelberg},\ \bibinfo
  {year} {2012})\BibitemShut {NoStop}%
\bibitem [{\citenamefont {Annett}(2004)}]{annett2004superconductivity}%
  \BibitemOpen
  \bibfield  {author} {\bibinfo {author} {\bibfnamefont {J.~F.}\ \bibnamefont
  {Annett}},\ }\href
  {https://global.oup.com/academic/product/superconductivity-superfluids-and-condensates-9780198507567?cc=us&lang=en&}
  {\emph {\bibinfo {title} {Superconductivity, superfluids and condensates}}},\
  Vol.~\bibinfo {volume} {5}\ (\bibinfo  {publisher} {Oxford University
  Press},\ \bibinfo {address} {Oxford},\ \bibinfo {year} {2004})\BibitemShut
  {NoStop}%
\bibitem [{\citenamefont {McDonald}(1995)}]{mcdonald1995cosmological}%
  \BibitemOpen
  \bibfield  {author} {\bibinfo {author} {\bibfnamefont {J.}~\bibnamefont
  {McDonald}},\ }\href {\doibase 10.1016/0370-2693(95)00716-X} {\bibfield
  {journal} {\bibinfo  {journal} {Phys. Lett. B}\ }\textbf {\bibinfo {volume}
  {357}},\ \bibinfo {pages} {19} (\bibinfo {year} {1995})}\BibitemShut
  {NoStop}%
\bibitem [{\citenamefont {Greiner}\ and\ \citenamefont
  {Reinhardt}(1996)}]{greiner1996field}%
  \BibitemOpen
  \bibfield  {author} {\bibinfo {author} {\bibfnamefont {W.}~\bibnamefont
  {Greiner}}\ and\ \bibinfo {author} {\bibfnamefont {J.}~\bibnamefont
  {Reinhardt}},\ }\href
  {https://link.springer.com/book/10.1007/978-3-642-61485-9#about-this-book}
  {\emph {\bibinfo {title} {Field Quantization}}}\ (\bibinfo  {publisher}
  {Springer-Verlag},\ \bibinfo {address} {Berlin, Heidelberg},\ \bibinfo {year}
  {1996})\BibitemShut {NoStop}%
\bibitem [{\citenamefont {Quevillon}\ \emph {et~al.}(2022)\citenamefont
  {Quevillon}, \citenamefont {Smith},\ and\ \citenamefont
  {Vuong}}]{quevillon2022axion}%
  \BibitemOpen
  \bibfield  {author} {\bibinfo {author} {\bibfnamefont {J.}~\bibnamefont
  {Quevillon}}, \bibinfo {author} {\bibfnamefont {C.}~\bibnamefont {Smith}}, \
  and\ \bibinfo {author} {\bibfnamefont {P.~N.~H.}\ \bibnamefont {Vuong}},\
  }\href {\doibase 10.1007/JHEP08(2022)137} {\bibfield  {journal} {\bibinfo
  {journal} {J. High Energ. Phys.}\ }\textbf {\bibinfo {volume} {2022}},\
  \bibinfo {pages} {137} (\bibinfo {year} {2022})}\BibitemShut {NoStop}%
\bibitem [{\citenamefont {Coleman}(2015)}]{coleman2015introduction}%
  \BibitemOpen
  \bibfield  {author} {\bibinfo {author} {\bibfnamefont {P.}~\bibnamefont
  {Coleman}},\ }\href
  {https://www.cambridge.org/core/books/introduction-to-manybody-physics/B7598FC1FCEE0285F5EC767E835854C8}
  {\emph {\bibinfo {title} {Introduction to many-body physics}}}\ (\bibinfo
  {publisher} {Cambridge University Press},\ \bibinfo {address} {Cambridge},\
  \bibinfo {year} {2015})\BibitemShut {NoStop}%
\bibitem [{\citenamefont {Tinkham}(1996)}]{tinkham1996introduction}%
  \BibitemOpen
  \bibfield  {author} {\bibinfo {author} {\bibfnamefont {M.}~\bibnamefont
  {Tinkham}},\ }\href {https://store.doverpublications.com/0486435032.html}
  {\emph {\bibinfo {title} {Introduction to superconductivity}}}\ (\bibinfo
  {publisher} {Dover Publications},\ \bibinfo {address} {New York},\ \bibinfo
  {year} {1996})\BibitemShut {NoStop}%
\bibitem [{\citenamefont {Mohapatra}(2022)}]{mohapatra2022baryogenesis}%
  \BibitemOpen
  \bibfield  {author} {\bibinfo {author} {\bibfnamefont {R.~N.}\ \bibnamefont
  {Mohapatra}},\ }\href
  {https://inspirehep.net/files/fe6e868dd450c98700c1e55e2c675eaa} {\bibfield
  {journal} {\bibinfo  {journal} {Acta Phys. Pol. B Proc. Suppl.}\ }\textbf
  {\bibinfo {volume} {15}},\ \bibinfo {pages} {2} (\bibinfo {year}
  {2022})}\BibitemShut {NoStop}%
\bibitem [{\citenamefont {Cline}\ \emph
  {et~al.}(2020{\natexlab{a}})\citenamefont {Cline}, \citenamefont {Puel},\
  and\ \citenamefont {Toma}}]{cline2020affleck}%
  \BibitemOpen
  \bibfield  {author} {\bibinfo {author} {\bibfnamefont {J.~M.}\ \bibnamefont
  {Cline}}, \bibinfo {author} {\bibfnamefont {M.}~\bibnamefont {Puel}}, \ and\
  \bibinfo {author} {\bibfnamefont {T.}~\bibnamefont {Toma}},\ }\href {\doibase
  10.1103/PhysRevD.101.043014} {\bibfield  {journal} {\bibinfo  {journal}
  {Phys. Rev. D}\ }\textbf {\bibinfo {volume} {101}},\ \bibinfo {pages}
  {043014} (\bibinfo {year} {2020}{\natexlab{a}})}\BibitemShut {NoStop}%
\bibitem [{\citenamefont {Cline}\ \emph
  {et~al.}(2020{\natexlab{b}})\citenamefont {Cline}, \citenamefont {Puel},\
  and\ \citenamefont {Toma}}]{cline2020little}%
  \BibitemOpen
  \bibfield  {author} {\bibinfo {author} {\bibfnamefont {J.}~\bibnamefont
  {Cline}}, \bibinfo {author} {\bibfnamefont {M.}~\bibnamefont {Puel}}, \ and\
  \bibinfo {author} {\bibfnamefont {T.}~\bibnamefont {Toma}},\ }\href {\doibase
  10.1007/JHEP05(2020)039} {\bibfield  {journal} {\bibinfo  {journal} {J. High
  Energ. Phys.}\ }\textbf {\bibinfo {volume} {05}},\ \bibinfo {pages} {039}
  (\bibinfo {year} {2020}{\natexlab{b}})}\BibitemShut {NoStop}%
\bibitem [{\citenamefont {Lloyd-Stubbs}\ and\ \citenamefont
  {McDonald}(2023)}]{lloyd2023leptogenesis}%
  \BibitemOpen
  \bibfield  {author} {\bibinfo {author} {\bibfnamefont {K.}~\bibnamefont
  {Lloyd-Stubbs}}\ and\ \bibinfo {author} {\bibfnamefont {J.}~\bibnamefont
  {McDonald}},\ }\href {\doibase 10.1103/PhysRevD.107.103511} {\bibfield
  {journal} {\bibinfo  {journal} {Phys. Rev. D}\ }\textbf {\bibinfo {volume}
  {107}},\ \bibinfo {pages} {103511} (\bibinfo {year} {2023})}\BibitemShut
  {NoStop}%
\bibitem [{\citenamefont {Haber}\ and\ \citenamefont
  {Surujon}(2012)}]{haber2012group}%
  \BibitemOpen
  \bibfield  {author} {\bibinfo {author} {\bibfnamefont {H.~E.}\ \bibnamefont
  {Haber}}\ and\ \bibinfo {author} {\bibfnamefont {Z.}~\bibnamefont
  {Surujon}},\ }\href {\doibase 10.1103/PhysRevD.86.075007} {\bibfield
  {journal} {\bibinfo  {journal} {Phys. Rev. D}\ }\textbf {\bibinfo {volume}
  {86}},\ \bibinfo {pages} {075007} (\bibinfo {year} {2012})}\BibitemShut
  {NoStop}%
\bibitem [{\citenamefont {Branco}\ and\ \citenamefont
  {Ivanov}(2016)}]{branco2016group}%
  \BibitemOpen
  \bibfield  {author} {\bibinfo {author} {\bibfnamefont {G.~C.}\ \bibnamefont
  {Branco}}\ and\ \bibinfo {author} {\bibfnamefont {I.~P.}\ \bibnamefont
  {Ivanov}},\ }\href {\doibase 10.1007/JHEP01(2016)116} {\bibfield  {journal}
  {\bibinfo  {journal} {J. High Energ. Phys.}\ }\textbf {\bibinfo {volume}
  {2016}},\ \bibinfo {pages} {116} (\bibinfo {year} {2016})}\BibitemShut
  {NoStop}%
\bibitem [{\citenamefont {Branco}\ \emph {et~al.}(1999)\citenamefont {Branco},
  \citenamefont {Lavoura},\ and\ \citenamefont {Silva}}]{branco1999cp}%
  \BibitemOpen
  \bibfield  {author} {\bibinfo {author} {\bibfnamefont {G.~C.}\ \bibnamefont
  {Branco}}, \bibinfo {author} {\bibfnamefont {L.}~\bibnamefont {Lavoura}}, \
  and\ \bibinfo {author} {\bibfnamefont {J.~P.}\ \bibnamefont {Silva}},\ }\href
  {\doibase
  https://global.oup.com/academic/product/cp-violation-9780198503996?cc=mo&lang=en&#}
  {\emph {\bibinfo {title} {CP violation}}}\ (\bibinfo  {publisher} {Oxford
  University Press},\ \bibinfo {address} {Oxford},\ \bibinfo {year}
  {1999})\BibitemShut {NoStop}%
\bibitem [{\citenamefont {Sozzi}(2007)}]{sozzi2007discrete}%
  \BibitemOpen
  \bibfield  {author} {\bibinfo {author} {\bibfnamefont {M.}~\bibnamefont
  {Sozzi}},\ }\href {\doibase 10.1093/acprof:oso/9780199296668.001.0001} {\emph
  {\bibinfo {title} {Discrete symmetries and CP violation: From experiment to
  theory}}}\ (\bibinfo  {publisher} {Oxford University Press},\ \bibinfo
  {address} {Oxford},\ \bibinfo {year} {2007})\BibitemShut {NoStop}%
\bibitem [{big(2009)}]{bigi2009cp}%
  \BibitemOpen
  \href {\doibase 10.1017/CBO9780511581014} {\emph {\bibinfo {title} {CP
  violation}}}\ (\bibinfo  {publisher} {Cambridge University Press},\ \bibinfo
  {address} {Cambridge},\ \bibinfo {year} {2009})\BibitemShut {NoStop}%
\bibitem [{\citenamefont {Cheung}\ \emph {et~al.}(2012)\citenamefont {Cheung},
  \citenamefont {Dahlen},\ and\ \citenamefont {Elor}}]{cheung2012bubble}%
  \BibitemOpen
  \bibfield  {author} {\bibinfo {author} {\bibfnamefont {C.}~\bibnamefont
  {Cheung}}, \bibinfo {author} {\bibfnamefont {A.}~\bibnamefont {Dahlen}}, \
  and\ \bibinfo {author} {\bibfnamefont {G.}~\bibnamefont {Elor}},\ }\href
  {\doibase 10.1007/JHEP09(2012)073} {\bibfield  {journal} {\bibinfo  {journal}
  {J. High Energ. Phys.}\ }\textbf {\bibinfo {volume} {2012}},\ \bibinfo
  {pages} {073} (\bibinfo {year} {2012})}\BibitemShut {NoStop}%
\bibitem [{\citenamefont {Comelli}\ \emph {et~al.}(1994)\citenamefont
  {Comelli}, \citenamefont {Pietroni},\ and\ \citenamefont
  {Riotto}}]{comelli1994spontaneous}%
  \BibitemOpen
  \bibfield  {author} {\bibinfo {author} {\bibfnamefont {D.}~\bibnamefont
  {Comelli}}, \bibinfo {author} {\bibfnamefont {M.}~\bibnamefont {Pietroni}}, \
  and\ \bibinfo {author} {\bibfnamefont {A.}~\bibnamefont {Riotto}},\ }\href
  {\doibase 10.1016/0550-3213(94)90511-8} {\bibfield  {journal} {\bibinfo
  {journal} {Nucl. Phys. B}\ }\textbf {\bibinfo {volume} {412}},\ \bibinfo
  {pages} {441} (\bibinfo {year} {1994})}\BibitemShut {NoStop}%
\bibitem [{\citenamefont {Coito}\ \emph {et~al.}(2021)\citenamefont {Coito},
  \citenamefont {Faubel}, \citenamefont {Herrero-Garcia},\ and\ \citenamefont
  {Santamaria}}]{coito2021dark}%
  \BibitemOpen
  \bibfield  {author} {\bibinfo {author} {\bibfnamefont {L.}~\bibnamefont
  {Coito}}, \bibinfo {author} {\bibfnamefont {C.}~\bibnamefont {Faubel}},
  \bibinfo {author} {\bibfnamefont {J.}~\bibnamefont {Herrero-Garcia}}, \ and\
  \bibinfo {author} {\bibfnamefont {A.}~\bibnamefont {Santamaria}},\ }\href
  {\doibase 10.1007/JHEP11(2021)202} {\bibfield  {journal} {\bibinfo  {journal}
  {J. High Energ. Phys.}\ }\textbf {\bibinfo {volume} {2021}},\ \bibinfo
  {pages} {202} (\bibinfo {year} {2021})}\BibitemShut {NoStop}%
\bibitem [{\citenamefont {Nambu}\ and\ \citenamefont
  {Jona-Lasinio}(1961)}]{nambu1961dynamical}%
  \BibitemOpen
  \bibfield  {author} {\bibinfo {author} {\bibfnamefont {Y.}~\bibnamefont
  {Nambu}}\ and\ \bibinfo {author} {\bibfnamefont {G.}~\bibnamefont
  {Jona-Lasinio}},\ }\href {\doibase 10.1103/PhysRev.122.345} {\bibfield
  {journal} {\bibinfo  {journal} {Phys. Rev.}\ }\textbf {\bibinfo {volume}
  {122}},\ \bibinfo {pages} {345} (\bibinfo {year} {1961})}\BibitemShut
  {NoStop}%
\bibitem [{\citenamefont {Goldstone}(1961)}]{goldstone1961field}%
  \BibitemOpen
  \bibfield  {author} {\bibinfo {author} {\bibfnamefont {J.}~\bibnamefont
  {Goldstone}},\ }\href {\doibase 10.1007/BF02812722} {\bibfield  {journal}
  {\bibinfo  {journal} {Nuovo Cimento}\ }\textbf {\bibinfo {volume} {19}},\
  \bibinfo {pages} {154} (\bibinfo {year} {1961})}\BibitemShut {NoStop}%
\bibitem [{\citenamefont {Goldstone}\ \emph {et~al.}(1962)\citenamefont
  {Goldstone}, \citenamefont {Salam},\ and\ \citenamefont
  {Weinberg}}]{goldstone1962broken}%
  \BibitemOpen
  \bibfield  {author} {\bibinfo {author} {\bibfnamefont {J.}~\bibnamefont
  {Goldstone}}, \bibinfo {author} {\bibfnamefont {A.}~\bibnamefont {Salam}}, \
  and\ \bibinfo {author} {\bibfnamefont {S.}~\bibnamefont {Weinberg}},\ }\href
  {\doibase 10.1103/PhysRev.127.965} {\bibfield  {journal} {\bibinfo  {journal}
  {Phys. Rev.}\ }\textbf {\bibinfo {volume} {127}},\ \bibinfo {pages} {965}
  (\bibinfo {year} {1962})}\BibitemShut {NoStop}%
\bibitem [{\citenamefont {Aad}\ \emph {et~al.}(2020)\citenamefont {Aad},
  \citenamefont {Abbott}, \citenamefont {Abbott}, \citenamefont {Abud},
  \citenamefont {Abeling}, \citenamefont {Abhayasinghe}, \citenamefont {Abidi},
  \citenamefont {AbouZeid}, \citenamefont {Abraham}, \citenamefont {Abramowicz}
  \emph {et~al.}}]{aad2020cp}%
  \BibitemOpen
  \bibfield  {author} {\bibinfo {author} {\bibfnamefont {G.}~\bibnamefont
  {Aad}}, \bibinfo {author} {\bibfnamefont {B.}~\bibnamefont {Abbott}},
  \bibinfo {author} {\bibfnamefont {D.~C.}\ \bibnamefont {Abbott}}, \bibinfo
  {author} {\bibfnamefont {A.~A.}\ \bibnamefont {Abud}}, \bibinfo {author}
  {\bibfnamefont {K.}~\bibnamefont {Abeling}}, \bibinfo {author} {\bibfnamefont
  {D.~K.}\ \bibnamefont {Abhayasinghe}}, \bibinfo {author} {\bibfnamefont
  {S.~H.}\ \bibnamefont {Abidi}}, \bibinfo {author} {\bibfnamefont {O.~S.}\
  \bibnamefont {AbouZeid}}, \bibinfo {author} {\bibfnamefont {N.~L.}\
  \bibnamefont {Abraham}}, \bibinfo {author} {\bibfnamefont {H.}~\bibnamefont
  {Abramowicz}},  \emph {et~al.},\ }\href {\doibase
  10.1103/PhysRevLett.125.061802} {\bibfield  {journal} {\bibinfo  {journal}
  {Phys. Rev. Lett.}\ }\textbf {\bibinfo {volume} {125}},\ \bibinfo {pages}
  {061802} (\bibinfo {year} {2020})}\BibitemShut {NoStop}%
\bibitem [{\citenamefont {Sirunyan}\ \emph {et~al.}(2020)\citenamefont
  {Sirunyan}, \citenamefont {Tumasyan}, \citenamefont {Adam}, \citenamefont
  {Ambrogi}, \citenamefont {Bergauer}, \citenamefont {Dragicevic},
  \citenamefont {Er{\"o}}, \citenamefont {Del~Valle}, \citenamefont {Flechl},
  \citenamefont {Fruehwirth} \emph {et~al.}}]{sirunyan2020measurements}%
  \BibitemOpen
  \bibfield  {author} {\bibinfo {author} {\bibfnamefont {A.~M.}\ \bibnamefont
  {Sirunyan}}, \bibinfo {author} {\bibfnamefont {A.}~\bibnamefont {Tumasyan}},
  \bibinfo {author} {\bibfnamefont {W.}~\bibnamefont {Adam}}, \bibinfo {author}
  {\bibfnamefont {F.}~\bibnamefont {Ambrogi}}, \bibinfo {author} {\bibfnamefont
  {T.}~\bibnamefont {Bergauer}}, \bibinfo {author} {\bibfnamefont
  {M.}~\bibnamefont {Dragicevic}}, \bibinfo {author} {\bibfnamefont
  {J.}~\bibnamefont {Er{\"o}}}, \bibinfo {author} {\bibfnamefont {A.~E.}\
  \bibnamefont {Del~Valle}}, \bibinfo {author} {\bibfnamefont {M.}~\bibnamefont
  {Flechl}}, \bibinfo {author} {\bibfnamefont {R.}~\bibnamefont {Fruehwirth}},
  \emph {et~al.},\ }\href {\doibase 10.1103/PhysRevLett.125.061801} {\bibfield
  {journal} {\bibinfo  {journal} {Phys. Rev. Lett.}\ }\textbf {\bibinfo
  {volume} {125}},\ \bibinfo {pages} {061801} (\bibinfo {year}
  {2020})}\BibitemShut {NoStop}%
\bibitem [{\citenamefont {Dvali}\ \emph {et~al.}(1995)\citenamefont {Dvali},
  \citenamefont {Tavartkiladze},\ and\ \citenamefont
  {Nanobashvili}}]{dvali1995biased}%
  \BibitemOpen
  \bibfield  {author} {\bibinfo {author} {\bibfnamefont {G.}~\bibnamefont
  {Dvali}}, \bibinfo {author} {\bibfnamefont {Z.}~\bibnamefont
  {Tavartkiladze}}, \ and\ \bibinfo {author} {\bibfnamefont {J.}~\bibnamefont
  {Nanobashvili}},\ }\href {\doibase 10.1016/0370-2693(95)00511-I} {\bibfield
  {journal} {\bibinfo  {journal} {Phys. Lett. B}\ }\textbf {\bibinfo {volume}
  {352}},\ \bibinfo {pages} {214} (\bibinfo {year} {1995})}\BibitemShut
  {NoStop}%
\bibitem [{\citenamefont {Gleiser}\ \emph {et~al.}(2008)\citenamefont
  {Gleiser}, \citenamefont {Rogers},\ and\ \citenamefont
  {Thorarinson}}]{gleiser2008bubbling}%
  \BibitemOpen
  \bibfield  {author} {\bibinfo {author} {\bibfnamefont {M.}~\bibnamefont
  {Gleiser}}, \bibinfo {author} {\bibfnamefont {B.}~\bibnamefont {Rogers}}, \
  and\ \bibinfo {author} {\bibfnamefont {J.}~\bibnamefont {Thorarinson}},\
  }\href {\doibase 10.1103/PhysRevD.77.023513} {\bibfield  {journal} {\bibinfo
  {journal} {Phys. Rev. D}\ }\textbf {\bibinfo {volume} {77}},\ \bibinfo
  {pages} {023513} (\bibinfo {year} {2008})}\BibitemShut {NoStop}%
\bibitem [{\citenamefont {Coleman}(1977)}]{coleman1977fate}%
  \BibitemOpen
  \bibfield  {author} {\bibinfo {author} {\bibfnamefont {S.}~\bibnamefont
  {Coleman}},\ }\href {\doibase 10.1103/PhysRevD.15.2929} {\bibfield  {journal}
  {\bibinfo  {journal} {Phys. Rev. D}\ }\textbf {\bibinfo {volume} {15}},\
  \bibinfo {pages} {2929} (\bibinfo {year} {1977})}\BibitemShut {NoStop}%
\bibitem [{\citenamefont {\textbf{Erratum}:
  S.~Coleman}(1977)}]{coleman1977erratum}%
  \BibitemOpen
  \bibfield  {author} {\bibinfo {author} {\bibnamefont {\textbf{Erratum}:
  S.~Coleman}},\ }\href {\doibase 10.1103/PhysRevD.16.1248} {\bibfield
  {journal} {\bibinfo  {journal} {Phys. Rev. D}\ }\textbf {\bibinfo {volume}
  {16}},\ \bibinfo {pages} {1248} (\bibinfo {year} {1977})}\BibitemShut
  {NoStop}%
\bibitem [{\citenamefont {Callan~Jr}\ and\ \citenamefont
  {Coleman}(1977)}]{callan1977fate}%
  \BibitemOpen
  \bibfield  {author} {\bibinfo {author} {\bibfnamefont {C.~G.}\ \bibnamefont
  {Callan~Jr}}\ and\ \bibinfo {author} {\bibfnamefont {S.}~\bibnamefont
  {Coleman}},\ }\href {\doibase 10.1103/PhysRevD.16.1762} {\bibfield  {journal}
  {\bibinfo  {journal} {Phys. Rev. D}\ }\textbf {\bibinfo {volume} {16}},\
  \bibinfo {pages} {1762} (\bibinfo {year} {1977})}\BibitemShut {NoStop}%
\bibitem [{\citenamefont {Kobsarev}\ \emph {et~al.}(1974)\citenamefont
  {Kobsarev}, \citenamefont {Voloshin},\ and\ \citenamefont
  {Okun}}]{kobsarev1974bubbles}%
  \BibitemOpen
  \bibfield  {author} {\bibinfo {author} {\bibfnamefont {I.~Y.}\ \bibnamefont
  {Kobsarev}}, \bibinfo {author} {\bibfnamefont {M.~B.}\ \bibnamefont
  {Voloshin}}, \ and\ \bibinfo {author} {\bibfnamefont {L.~B.}\ \bibnamefont
  {Okun}},\ }\href {\doibase
  https://inis.iaea.org/search/search.aspx?orig_q=RN:7276129} {\bibfield
  {journal} {\bibinfo  {journal} {Yad. Fiz.}\ }\textbf {\bibinfo {volume}
  {20}},\ \bibinfo {pages} {1229} (\bibinfo {year} {1974})},\ \bibinfo {note}
  {[Sov. J. Nucl. Phys. 20, 644 (1975)]}\BibitemShut {NoStop}%
\bibitem [{\citenamefont {Endo}\ \emph {et~al.}(2017)\citenamefont {Endo},
  \citenamefont {Moroi}, \citenamefont {Nojiri},\ and\ \citenamefont
  {Shoji}}]{endo2017false}%
  \BibitemOpen
  \bibfield  {author} {\bibinfo {author} {\bibfnamefont {M.}~\bibnamefont
  {Endo}}, \bibinfo {author} {\bibfnamefont {T.}~\bibnamefont {Moroi}},
  \bibinfo {author} {\bibfnamefont {M.~M.}\ \bibnamefont {Nojiri}}, \ and\
  \bibinfo {author} {\bibfnamefont {Y.}~\bibnamefont {Shoji}},\ }\href
  {\doibase 10.1007/JHEP11(2017)074} {\bibfield  {journal} {\bibinfo  {journal}
  {J. High Energ. Phys.}\ }\textbf {\bibinfo {volume} {2017}},\ \bibinfo
  {pages} {074} (\bibinfo {year} {2017})}\BibitemShut {NoStop}%
\bibitem [{\citenamefont {Chigusa}\ \emph {et~al.}(2020)\citenamefont
  {Chigusa}, \citenamefont {Moroi},\ and\ \citenamefont
  {Shoji}}]{chigusa2020precise}%
  \BibitemOpen
  \bibfield  {author} {\bibinfo {author} {\bibfnamefont {S.}~\bibnamefont
  {Chigusa}}, \bibinfo {author} {\bibfnamefont {T.}~\bibnamefont {Moroi}}, \
  and\ \bibinfo {author} {\bibfnamefont {Y.}~\bibnamefont {Shoji}},\ }\href
  {\doibase 10.1007/JHEP11(2020)006} {\bibfield  {journal} {\bibinfo  {journal}
  {J. High Energ. Phys.}\ }\textbf {\bibinfo {volume} {2020}},\ \bibinfo
  {pages} {006} (\bibinfo {year} {2020})}\BibitemShut {NoStop}%
\bibitem [{\citenamefont {Andreassen}\ \emph {et~al.}(2017)\citenamefont
  {Andreassen}, \citenamefont {Farhi}, \citenamefont {Frost},\ and\
  \citenamefont {Schwartz}}]{andreassen2017precision}%
  \BibitemOpen
  \bibfield  {author} {\bibinfo {author} {\bibfnamefont {A.}~\bibnamefont
  {Andreassen}}, \bibinfo {author} {\bibfnamefont {D.}~\bibnamefont {Farhi}},
  \bibinfo {author} {\bibfnamefont {W.}~\bibnamefont {Frost}}, \ and\ \bibinfo
  {author} {\bibfnamefont {M.~D.}\ \bibnamefont {Schwartz}},\ }\href {\doibase
  10.1103/PhysRevD.95.085011} {\bibfield  {journal} {\bibinfo  {journal} {Phys.
  Rev. D}\ }\textbf {\bibinfo {volume} {95}},\ \bibinfo {pages} {085011}
  (\bibinfo {year} {2017})}\BibitemShut {NoStop}%
\bibitem [{\citenamefont {Devoto}\ \emph {et~al.}(2022)\citenamefont {Devoto},
  \citenamefont {Devoto}, \citenamefont {Di~Luzio},\ and\ \citenamefont
  {Ridolfi}}]{devoto2022false}%
  \BibitemOpen
  \bibfield  {author} {\bibinfo {author} {\bibfnamefont {F.}~\bibnamefont
  {Devoto}}, \bibinfo {author} {\bibfnamefont {S.}~\bibnamefont {Devoto}},
  \bibinfo {author} {\bibfnamefont {L.}~\bibnamefont {Di~Luzio}}, \ and\
  \bibinfo {author} {\bibfnamefont {G.}~\bibnamefont {Ridolfi}},\ }\href
  {\doibase /10.1088/1361-6471/ac7f24} {\bibfield  {journal} {\bibinfo
  {journal} {J. Phys. G: Nucl. Part. Phys.}\ }\textbf {\bibinfo {volume}
  {49}},\ \bibinfo {pages} {103001} (\bibinfo {year} {2022})}\BibitemShut
  {NoStop}%
\bibitem [{\citenamefont {Coleman}\ and\ \citenamefont
  {De~Luccia}(1980)}]{coleman1980gravitational}%
  \BibitemOpen
  \bibfield  {author} {\bibinfo {author} {\bibfnamefont {S.}~\bibnamefont
  {Coleman}}\ and\ \bibinfo {author} {\bibfnamefont {F.}~\bibnamefont
  {De~Luccia}},\ }\href {\doibase 10.1103/PhysRevD.21.3305} {\bibfield
  {journal} {\bibinfo  {journal} {Phys. Rev. D}\ }\textbf {\bibinfo {volume}
  {21}},\ \bibinfo {pages} {3305} (\bibinfo {year} {1980})}\BibitemShut
  {NoStop}%
\bibitem [{\citenamefont {Andreassen}\ \emph {et~al.}(2016)\citenamefont
  {Andreassen}, \citenamefont {Farhi}, \citenamefont {Frost},\ and\
  \citenamefont {Schwartz}}]{andreassen2016direct}%
  \BibitemOpen
  \bibfield  {author} {\bibinfo {author} {\bibfnamefont {A.}~\bibnamefont
  {Andreassen}}, \bibinfo {author} {\bibfnamefont {D.}~\bibnamefont {Farhi}},
  \bibinfo {author} {\bibfnamefont {W.}~\bibnamefont {Frost}}, \ and\ \bibinfo
  {author} {\bibfnamefont {M.~D.}\ \bibnamefont {Schwartz}},\ }\href {\doibase
  10.1103/PhysRevLett.117.231601} {\bibfield  {journal} {\bibinfo  {journal}
  {Phys. Rev. Lett.}\ }\textbf {\bibinfo {volume} {117}},\ \bibinfo {pages}
  {231601} (\bibinfo {year} {2016})}\BibitemShut {NoStop}%
\bibitem [{\citenamefont {Maggiore}(2005)}]{maggiore2005modern}%
  \BibitemOpen
  \bibfield  {author} {\bibinfo {author} {\bibfnamefont {M.}~\bibnamefont
  {Maggiore}},\ }\href
  {https://global.oup.com/academic/product/a-modern-introduction-to-quantum-field-theory-9780198520733?cc=mo&lang=en&}
  {\emph {\bibinfo {title} {A Modern Introduction to Quantum Field Theory}}}\
  (\bibinfo  {publisher} {Oxford University Press},\ \bibinfo {address}
  {Oxford},\ \bibinfo {year} {2005})\BibitemShut {NoStop}%
\bibitem [{\citenamefont {Weinberg}\ and\ \citenamefont
  {Greenberg}(1996)}]{weinberg1996quantum}%
  \BibitemOpen
  \bibfield  {author} {\bibinfo {author} {\bibfnamefont {S.}~\bibnamefont
  {Weinberg}}\ and\ \bibinfo {author} {\bibfnamefont {O.~W.}\ \bibnamefont
  {Greenberg}},\ }\href {\doibase
  https://www.cambridge.org/mo/universitypress/subjects/physics/theoretical-physics-and-mathematical-physics/quantum-theory-fields-volume-2?format=HB&isbn=9780521550024}
  {\emph {\bibinfo {title} {The Quantum Theory of Fields, Volume 2: Modern
  Applications}}}\ (\bibinfo  {publisher} {Cambridge University Press},\
  \bibinfo {address} {Cambridge},\ \bibinfo {year} {1996})\BibitemShut
  {NoStop}%
\bibitem [{\citenamefont {Volovik}\ and\ \citenamefont
  {Zubkov}(2014)}]{volovik2014higgs}%
  \BibitemOpen
  \bibfield  {author} {\bibinfo {author} {\bibfnamefont {G.~E.}\ \bibnamefont
  {Volovik}}\ and\ \bibinfo {author} {\bibfnamefont {M.~A.}\ \bibnamefont
  {Zubkov}},\ }\href {\doibase 10.1007/s10909-013-0905-7} {\bibfield  {journal}
  {\bibinfo  {journal} {J. Low Temp. Phys.}\ }\textbf {\bibinfo {volume}
  {175}},\ \bibinfo {pages} {486} (\bibinfo {year} {2014})}\BibitemShut
  {NoStop}%
\bibitem [{\citenamefont {Agrawal}\ \emph {et~al.}(2020)\citenamefont
  {Agrawal}, \citenamefont {Saha}, \citenamefont {Xu}, \citenamefont {Yu},\
  and\ \citenamefont {Yuan}}]{agrawal2020determining}%
  \BibitemOpen
  \bibfield  {author} {\bibinfo {author} {\bibfnamefont {P.}~\bibnamefont
  {Agrawal}}, \bibinfo {author} {\bibfnamefont {D.}~\bibnamefont {Saha}},
  \bibinfo {author} {\bibfnamefont {L.}~\bibnamefont {Xu}}, \bibinfo {author}
  {\bibfnamefont {J.}~\bibnamefont {Yu}}, \ and\ \bibinfo {author}
  {\bibfnamefont {C.}~\bibnamefont {Yuan}},\ }\href {\doibase
  10.1103/PhysRevD.101.075023} {\bibfield  {journal} {\bibinfo  {journal}
  {Phys. Rev. D}\ }\textbf {\bibinfo {volume} {101}},\ \bibinfo {pages}
  {075023} (\bibinfo {year} {2020})}\BibitemShut {NoStop}%
\end{thebibliography}%


%
\end{document}